\documentclass[fleqn,usenatbib]{mnras}
\usepackage{newtxtext,newtxmath}
\usepackage[T1]{fontenc}
\usepackage{ae,aecompl}
\usepackage{bm}
\usepackage{times}

\usepackage{graphicx}
\usepackage{amsmath}
\usepackage{gensymb}

\hypersetup{pdfauthor={R. Hirai},
            pdftitle={Simulating the formation of Eta Carinae's surrounding nebula through unstable triple evolution and stellar merger-induced eruption},
            pdfkeywords={stars: individual: Eta Carinae -- binaries: close -- stars: winds, outflows -- stars: kinematics and dynamics},
            bookmarksnumbered=true}

\newcommand{\msun}{\textnormal{M}_\odot}
\newcommand{\rsun}{\textnormal{R}_\odot}
\newcommand{\etac}{$\eta$ Car}

\title[Formation of {\etac}'s surrounding nebula]{Simulating the formation of $\eta$ Carinae's surrounding nebula through unstable triple evolution and stellar merger-induced eruption}
\author[R. Hirai]{Ryosuke Hirai$^{1,2,3}$\thanks{E-mail: ryosuke.hirai@monash.edu}, Philipp Podsiadlowski$^{3,4}$, Stanley P. Owocki$^{5}$,
\newauthor Fabian R. N. Schneider$^{6,7,3}$, Nathan Smith$^{8}$
\\
$^{1}$OzGrav: Australian Research Council Centre of Excellence for Gravitational Wave Discovery, Clayton, VIC 3800, Australia\\
$^{2}$Monash Centre for Astrophysics, School of Physics and Astronomy, Monash University, Clayton, Victoria 3800, Australia\\
$^{3}$Department of Physics, University of Oxford, Keble Rd, Oxford, OX1 3RH, United Kingdom\\
$^{4}$Argelander-Institut f\"ur Astronomie der Universit\"at Bonn, Auf dem H\"ugel 71, D-53121 Bonn, Germany\\
$^{5}$Bartol Research Institute, Department of Physics \& Astronomy, University of Delaware, Newark, DE 19716 United States of America\\
$^{6}$Heidelberger Institut f\"{u}r Theoretische Studien, Schloss-Wolfsbrunnenweg 35, 69118 Heidelberg, Germany\\
$^{7}$Astronomisches Rechen-Institut, Zentrum f\"ur Astronomie der Universit\"at Heidelberg, M\"onchhofstr. 12-14, D-69120 Heidelberg, Germany\\
$^{8}$Steward Observatory, University of Arizona, 933 N. Cherry Ave., Tucson, AZ 85721, United States of America
}

\date{Accepted XXX. Received YYY; in original form ZZZ}

\pubyear{2021}
\begin{document}
\label{firstpage}
\pagerange{\pageref{firstpage}--\pageref{lastpage}}
\maketitle

\begin{abstract}
$\eta$ Carinae is an extraordinary massive star famous for its 19th century Great Eruption and the surrounding Homunculus nebula ejected in that event. The cause of this eruption has been the centre of a long-standing mystery. Recent observations, including light-echo spectra of the eruption, suggest that it most likely resulted from a stellar merger in an unstable triple system. Here we present a detailed set of theoretical calculations for this scenario; from the dynamics of unstable triple systems and the mass ejection from close binary encounters, to the mass outflow from the eruption caused by the stellar merger and the post-merger wind phase. In our model the bipolar post-merger wind is the primary agent for creating the Homunculus, as it sweeps up external eruption ejecta into a thin shell. Our simulations reproduce many of the key aspects of the shape and kinematics of both the Homunculus nebula and its complex surrounding structure, providing strong support for the merger-in-a-triple scenario.
\end{abstract}


\section{Introduction}
The remarkable bipolar Homunculus nebula surrounding Eta Carinae (\etac) has fascinated astronomers for decades. It has two lobes emanating from the central star, with a ``skirt''-like feature in the equatorial plane \citep[]{tha49,gav50,rin58,hac86,hil92,duschl95,mor98,dav01,smith02,smi06a,ste14}. The Homunculus nebula is usually associated with the ``Great Eruption'' that occurred in the 1840's when {\etac} became the second brightest star in the sky \citep{fre04}. A significant amount of mass was ejected in the Great Eruption, estimated to be $\ga10~\msun$ with an energy of $\sim10^{50}$~erg \citep[e.g.][]{smi03b,smi06a}. Most of the mass ejected in the mid 19th century is contained in the Homunculus shell, but faster components are known to exist outside, with much less mass but possibly comparable kinetic energy \citep[]{smi08,meh16,smi18a,smi19}. Peak brightness was recorded in December 1844, but there are known to be some precursor eruptions in 1838 and 1843 and also later lesser eruptions in 1890 and 1940 \citep[]{fre04,fer09,smi11a}. The times of precursor eruptions are consistent with the periastron passages of a current-day wide binary companion (explained later), while the Great and later eruptions seem to be unrelated \citep[]{smi11a}. The later eruption in 1890 led to the formation of the ``Little Homunculus'' \citep[]{ish03,smi05}.

Apart from all these mass ejections that were visible in the light curve, it is known that there were more prior ejection episodes well before the Great Eruption, before a good photometric record was available. The matter ejected in these historical ejections are located well outside the Homunculus nebula, known as the ``Outer Ejecta'', with velocities slower than that of the Homunculus shell \citep[]{tha50,wal76,wal78,smi08,kim16,meh16}. A rough estimate of the ejection dates can be made using the proper motion of the ejecta in {\it Hubble Space Telescope} ({\it{HST}}) images, and it seems that there were at least three distinct mass ejection episodes with $\sim300$~yr intervals \citep[]{kim16}. More recently ejected material have higher velocities and some of the inner ejectiles are overtaking the older slower ejectiles. Interestingly, these older ejections were not spherically symmetric or even axisymmetric. Each major historical ejection seems to have had a different orientation and a different opening angle, randomly oriented and unrelated to the symmetry axis of the Homunculus. Soft X-ray emission is also observed from the position of the Outer Ejecta \citep[]{sew79,sew01,wei04}, which is interpreted as emission from the fast ejecta from the Great Eruption running into slower previous ejecta \citep[]{sm04,smi08,meh16,smi19}. There are also structures observed in H$\beta$ that fill in the gap between the Outer Ejecta and the Homunculus in the form of a bent cylinder, known as the ``ghost shell'' and ``outer shell'' \citep[]{cur02,meh16}, while most of the volume between the Homunculus and Outer Ejecta is filled with low-density gas seen in Mg~{\sc ii} resonance scattering \citep{smi19}.

Today, {\etac} is emitting an extremely strong stellar wind, reaching mass-loss rates up to $\dot{M}\sim10^{-3}~\msun~\mathrm{yr}^{-1}$ and terminal velocities of $\sim$600--1000~km~s$^{-1}$ \citep[]{vio89,dam98,smi03a,hil06}. There is a latitudinal dependence on the wind strength, having stronger mass-loss rates and higher velocities towards the poles \citep[]{smi03a}. This is in good agreement with so-called gravity darkened wind models where stars with rapid rotation ($\gtrsim$70\% critical) have larger radiative fluxes around the poles compared to the equator and therefore have stronger radiative driving\footnote{Some studies suggest that the latitudinal line profile variation can be explained without invoking rapid rotation \citep[]{gro12}.} \citep[]{cra95,owo96,owo97,owo98,mae00}.

This strong wind is also known to be interacting with a binary companion that is orbiting {\etac} on a $\sim5.54$ yr period \citep[]{dam96,dam97,dam00}. The companion star drives a strong wind that collides with the primary wind, producing hard X-rays with strong variability \citep[]{cor95,cor97,ishi99,gul09,gul11}. By comparing the X-ray observations with 3D hydrodynamical modelling, properties of the massive binary have been fairly well constrained despite not being able to directly image the companion \citep[]{pit02,mad12,mad13,cle14,rus16,bus19}. For example, the wind momenta of the two components have to be comparable in order to have a strong enough wind-wind interaction, so the estimated wind parameters for the secondary are $\dot{M}\sim10^{-5}$~$\msun$~yr$^{-1}$ and $v_\infty\sim3000$~km~s$^{-1}$. The strong variability indicates that the orbit is highly eccentric, with estimated eccentricities of $e\sim0.9$ \citep[]{nie07,kas16,gra20}. Modelling of the X-ray light curve and spatially resolved [Fe \textsc{iii}] emission enables us to decipher the 3D orientation of the orbit. It suggests that the orbital plane is aligned to the Homunculus symmetry plane within $\sim10$~degrees \citep[]{mad12} and the apastron direction is coincident with some non-axisymmetric features of the surrounding nebula \citep[]{ste14,smi18c}.

Many attempts have been made to model the eruptive mass loss of this extraordinary star. Initial attempts involved steady super-Eddington winds driven by the high luminosity from luminous blue variables \citep[]{sha00,owo04,smi06b,vma08,har09,sha10,owo16,qua16,owo17}. These models showed that in extreme cases, the large radiative luminosity observed during the Great Eruption is capable of driving steady winds with strengths compatible with the inferred high mass-loss rate. However, it requires an additional energy source apart from the steady-state core nuclear burning and it is not clear how the $\sim$10$^{50}$ erg of extra energy is supplied. Also, because these are single-star models, it requires rapid rotation to produce a bipolar nebula and it is again not clear how the large amount of angular momentum can be provided, or how rapid rotaton can persist after such extreme mass loss.

In any case, the enhanced wind models predict the Great Eruption to have a more or less fixed velocity. However, recent observations of light echoes of the Great Eruption have revealed that there is a very fast velocity component in the ejecta ($v_\infty\sim$10,000--20,000~km~s$^{-1}$) that cannot be explained within this scenario \citep[]{smi18a,smi18b}.

Instead of rapid rotation, some models rely on the companion star for the shaping of the Homunculus \citep{sok01,sok04,sok07,kas10,aka16}. In these models, the matter from the Great Eruption is partly accreted onto the main-sequence companion through an accretion disk. Part of the accreted matter is then emitted as bipolar jets, providing poleward kinetic energy to the Great Eruption ejecta. More recent modelling shows that this model is capable of producing the fast velocity component too \citep[]{aka20}.

Another possible channel is through pulsational pair-instability events of very massive stars \citep[]{bar67,yos16,woo17,leu19}. Stars with initial masses of $M_\mathrm{ini}\gtrsim80\msun$\footnote{The exact mass range is very uncertain.} are known to create cores where the effects of electron-positron pair production significantly affects its structure. The reduction of pressure due to pair production leads to a dynamically unstable implosion, which in turn ignites runaway oxygen burning. In stars with $M_\mathrm{ini}\gtrsim140\msun$, the energy released by this process is large enough to completely expell the entire star as a supernova explosion. However, stars in the range $80\msun\lesssim M_\mathrm{ini}\lesssim140\msun$ generate much less energy and thus expel only a part of its envelope. The process iterates until the oxygen content is exhausted and ends up as a normal core-collapse supernova or a failed supernova. In terms of the mass and energy budget, \etac's Great Eruption and its subsequent lesser eruptions could have resulted from these pulsational pair-instability events, but whether they can produce the bipolar nebula, its alignment with an eccentric companion star, or the time-scale of repeating outbursts is again unclear.

On the other hand, the bipolar shape of the nebula and the explosive nature of the Great Eruption might be naturally expected in a binary stellar merger scenario \citep[]{gal89,ibe99,pod06,mor06,pod10,fit12,por16,smi18b,owo19}. When two massive stars merge, the energy that is released from the decay of the binary orbit is deposited in the merger product; the total energy released is roughly given by the orbital energy of the immersed binary at the stage when either the spiralling-in secondary or the core of the primary (or both) are being tidally torn apart; this energy is of the order of the core binding energy ($\sim10^{50}$~erg) and is comparable to that of the Great Eruption \citep{smi03b}. The angular momentum of the orbit defines a special direction that could relate to the bipolar axisymmetrical structure of the Homunculus \citep{sok04,mor06}. For example, \citet{mor06} simulated the outflow from the merger of a red supergiant with a main-sequence companion with a combined mass of $20~\msun$ and find a bipolar distribution of ejecta.

A major difficulty for a simple binary merger scenario is that it is expected to be a terminal event producing only one eruption. Additional mechanisms would be required to explain the other eruptions before (e.g. Outer Ejecta) and after (e.g. Little Homunculus) the Great Eruption. The existence of a companion star today may resolve part of this issue. If the Great Eruption was caused by a merger, it means that the original system must have been a triple system. The complicated evolution of unstable triple systems \citep[]{per12,sha13,mic14} have been suggested to cause grazing collisions that create the seemingly random distribution of the Outer Ejecta \citep[]{smi18b}.  Moreover, \citet{smi18b} proposed a specific merger-in-a-triple scenario wherein mass transfer in the inner binary led to an exchange of partners that ejected the original stripped primary star on an eccentric orbit (observed now as the current wide companion), while sending the original tertiary inward to merge with the mass gainer, thus causing the Great Eruption.

This paper systematically investigates this merger scenario through hydrodynamical simulations for the Great Eruption and the formation of the Homunculus, as well as dynamical models of the 3-body interactions that led to prior ejecta. In Section \ref{sec:framework} we outline the framework of the model we pursue in this paper. In Section \ref{sec:greateruption} we present results of hydrodynamical simulations of the merger phase and how it compares with observed features of the Homunculus nebula today. Then we discuss possible triple evolution scenarios that lead to a merger and how it can create the Outer Ejecta in Section \ref{sec:sprays}. We speculate on the post-merger evolution of the merger product in Section \ref{sec:postmerger}, and discuss the origin of other observed features in Section \ref{sec:other}. We summarize our results and briefly discuss applications to other astrophysical phenomena in Section \ref{sec:summary}.

\section{Framework}\label{sec:framework}

\begin{figure*}
 \centering
 \includegraphics[width=0.6\linewidth]{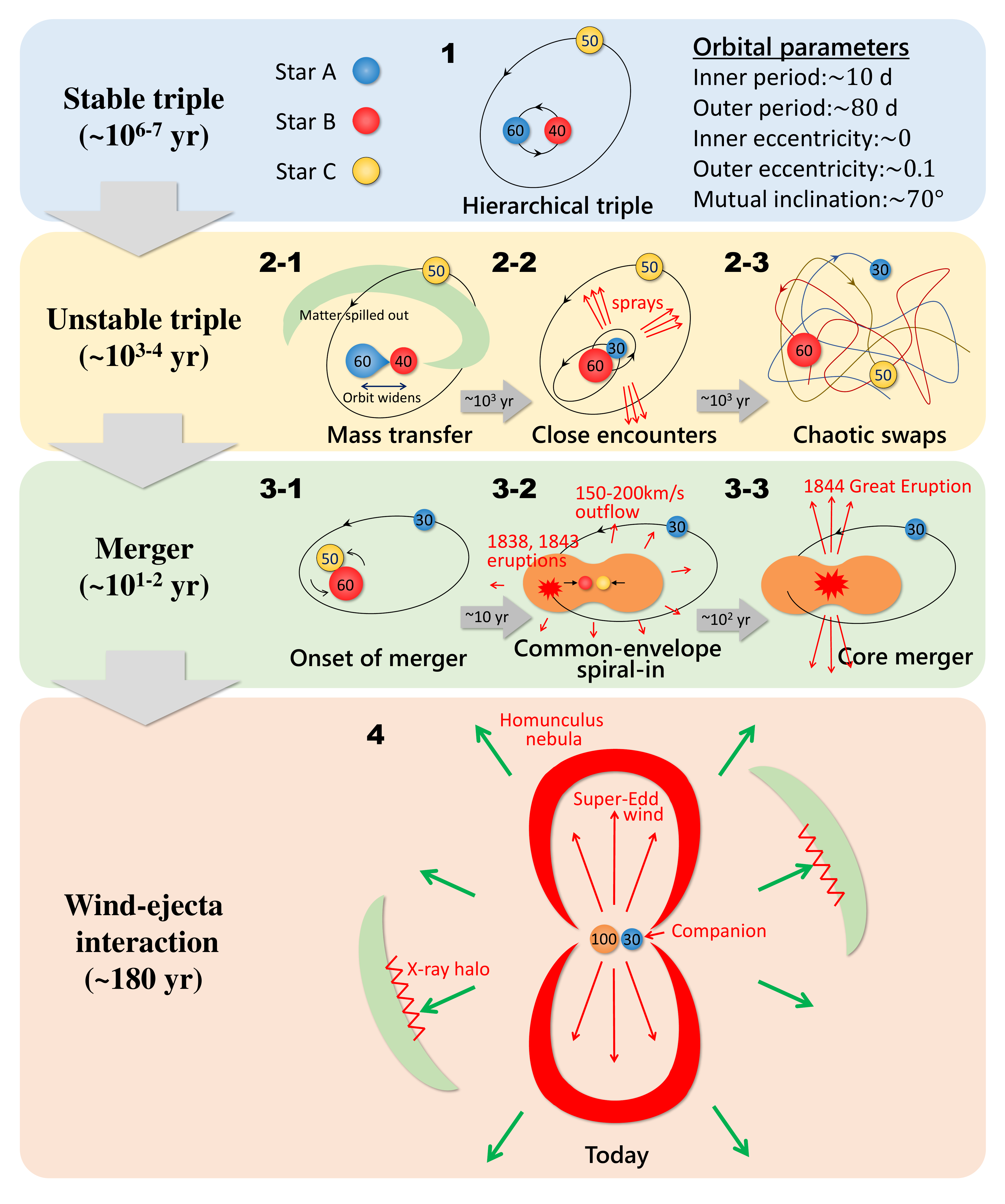}
 \caption{Schematic picture of the scenario for the formation of {\etac} that is investigated in this paper. Features illustrated in red are related to direct observables. Colours of stars are simply labels and do not express the physical colours.\label{fig:framework}}
\end{figure*}

Here we outline the framework of the scenario that we pursue in this paper.
The scenario combines previously proposed models for the triple evolution and merger \citep{smi18b} and shaping of the Homunculus \citep{owo05,mor06} in four phases as depicted in Figure~\ref{fig:framework}.

The evolution starts off with three massive stars in a hierarchical triple system (Phase 1). The masses of the stars are all similar and the mutual inclination of the inner and outer orbit is high enough to induce so-called Kozai-Lidov oscillations. Kozai-Lidov oscillations are a dynamical phenomenon in hierarchical triple systems where the eccentricity and inclination of the inner orbit exchange their values over secular time-scales\footnote{It was recently suggested that von Zeipel established the theoretical framework of the Kozai-Lidov mechanism more than 50 years before Kozai and Lidov did in the 1960s \citep[]{zei10,ito19}. We nevertheless use the conventional name in this paper.} \citep[]{zei10,koz62,lid62}. The distance between the two inner stars do not get close enough even at the peak eccentricities reached in the Kozai-Lidov cycles, so the stars go through their standard main-sequence evolution without interacting. 

Once the initially most massive star ends its main-sequence phase, the envelope expands and starts transferring matter to its companion (Phase 2-1). The mass transfer preferentially occurs when the orbit is most eccentric during the Kozai-Lidov cycles. Until the mass ratio inverts, the mass transfer is likely non-conservative and the matter spilled out of the system would be shaped like a partial torus and located on the orbital plane at eccentricity peaks. After the mass ratio inverts, the orbit widens with mass transfer due to angular momentum conservation. When the primary star has lost most of its hydrogen-rich envelope, it starts blowing a strong stellar wind. Because Kozai-Lidov oscillation time-scales are proportional to the period ratio of the inner and outer orbits, the Kozai-Lidov time-scale gradually shortens as the inner orbit widens over the mass transfer time-scale of about $\sim$10$^{4-5}$~yr. Once the period ratio becomes sufficiently small, the system is no longer stable and becomes mildly chaotic. In these so-called quasi-secular regimes where the orbit is chaotic but still quasi-periodic \citep[e.g.][]{ant12,sha13,ant14,mic14}, the eccentricity of the inner orbit can sometimes become large enough that the two stars almost touch each others surfaces at periastron (Phase 2-2). These grazing encounters can unbind part of the surface material and send them out in confined directions, which become the Outer Ejecta \citep[]{kim16}. The stochastic encounters between the stars eventually destabilize the orbit up to a point where the hierarchy of the orbits is completely lost and enters a chaotic phase (Phase 2-3). Such systems are very unstable and the stars can experience very close encounters.

When the two larger (in radius) stars approach at close distances, the envelopes will crash into each other and rapidly dissipate their orbital energies (Phase 3-1). This develops a brief common-envelope phase where the cores of the stars orbit inside the hydrogen envelope while the tertiary star orbits around the envelope on a stable eccentric orbit. The envelope is spun up rapidly and becomes extremely oblate because of the orbital angular momentum brought in, so the tertiary star can plunge into the bloated envelope at each periastron passage which can create some transient phenomena (Phase 3-2). Such transients may be related to the precursor eruptions seen in 1838 and 1843 \citep[]{smi11a}.

The frictional force acting on the cores will transfer energy from the orbit to the envelope and cause the orbit to shrink gradually. As this spiral-in time-scale becomes comparable to the orbital time-scale, the cores will rapidly approach each other, leading to a tidal disruption or a direct collision of the cores. This releases a substantial amount of energy and angular momentum at the centre of the oblate envelope on a time-scale of the order of the orbital period. Because this is much shorter than the dynamical time-scale of the envelope, it inevitably steepens into an outgoing shock, eventually reaching the surface and resulting in an explosion that is observed as the Great Eruption (Phase 3-3). It is easier for the shock to escape through the poles than the equator because of the oblateness of the envelope, so it naturally creates a bipolar explosion \citep[]{mor06}.

After the Great Eruption, the merger product still contains a large excess of energy and angular momentum. This excess energy enables the star to develop extremely strong super-Eddington winds. The wind sweeps up the inner parts of the ejecta, and the high density enables it to quickly radiatively cool into a thin shell (Phase 4). Because of the residual angular momentum, the merger product is rapidly rotating. Thus it has lower net effective gravity near the equator, with an associated ``gravity darkening'' \citep[]{zei24}. This makes the stellar wind weaker and slower from the equator, faster and denser over the poles. The bipolar wind blowing into a bipolar ejecta will create a hollow bipolar shell, which is what is observed as the Homunculus nebula today.

As the material is swept up into a dense cool shell, it creates an ideal situation for dust condensation. This only occurs after the shell has sufficiently expanded ($\gtrsim$1000~AU), where the shell cools down below the dust condensation temperature ($\lesssim$1500~K). As dust is formed, the opacity increases in the shell and radiation from the inner star can impart part of its momentum to the dust grains. This can in principle further accelerate the Homunculus shell, but the expected effect is negligible \citep[$\lesssim$1~km~s$^{-1}$; but see also][]{gla18}.

Not all of the material from the Great Eruption is swept up by the wind yet. There is some matter outside the Homunculus shell expanding outwards faster than the shell velocity but with a lower density and opacity that makes it more difficult to observe. This fast material can produce X-rays when it catches up with slower pre-eruption ejecta. If it catches up with the matter ejected from the close encounters during the triple evolution, the X-rays will be emitted from roughly the same place as the outer ejectiles. If it interacts with matter spilled out from the mass-transfer phase (panel 2 in Figure~\ref{fig:framework}), it will be emitted from a partial ring-like region. It could have also interacted with pre-merger wind material. The current images from X-ray telescopes are roughly consistent with both scenarios \citep[]{sew01}.

\section{The Great Eruption and formation of the Homunculus nebula}\label{sec:greateruption}

In this section we present our hydrodynamical simulation of the mass outflow from the stellar merger.

\subsection{Eruption simulation}\label{sec:eruption}
In order to mimic a merger product of two massive stars with a combined mass of $\sim100~\msun$, we first create a $100~\msun$ main-sequence star model using the public stellar evolution code \textsc{mesa} \citep[v10398;][]{MESA1,MESA2,MESA3,MESA4,MESA5}. Assuming a metallicity of $Z=0.02$\footnote{The exact value of metallicity does not influence our hydrodynamical simulations.}, we evolve the star up to the point where the central H mass abundance becomes lower than 0.2. This assumes that the merging stars are $\sim$80~per~cent into its main-sequence lifetime by the time the primary star finished its main sequence and the system became unstable. The same stellar model was used in our preceding study \citep[]{owo19}. We then map this star onto the centre of a spherical grid of our hydrodynamical code \textsc{hormone} \citep[]{RH16}, assuming axisymmetry and equatorial symmetry. See Appendix \ref{app:code} for the notation of variables, basic equations and details of the numerics. A stellar wind model is attached as a background with a mass-loss rate of $\dot{M}=10^{-5}~\msun$~yr$^{-1}$ and terminal velocity $v_\infty=400$~km~s$^{-1}$. The mass and momentum in this wind is tiny compared to the eruption ejecta, so the dynamics of the outflow is insensitive to the choice of our pre-eruption wind parameters. 

We follow a similar procedure as in \citet{mor06} but modified to match the current scenario to follow the dynamics of the merger. In particular, we focus on constructing a method so that the total energy and angular momentum in the system is consistent with what is available in the system. To the star, we apply a fixed spin-up rate $\dot{\Omega}_\mathrm{add}\sim1.6\times10^{-10}$~rad\;s$^{-2}$ to all cells where the angular velocity is sub-Keplerian $v_\varphi^2<|\phi|$ ($\phi$ is the gravitational potential). The spin-up rate is chosen so that the star will be slowly spun up to critical rotation over at least $\sim5$ dynamical time-scales and it can adjust its structure in a quasi-stationary manner. This procedure assumes that the angular momentum injected from the orbit into the envelope can quickly redistribute into a rigidly rotating core with a Keplerian envelope, which is often seen as end products of merger simulations of stellar and compact objects \citep[e.g.][]{ji13,fuj18,sch19}. The spin-up is ceased once a satisfactory amount of total angular momentum is injected, and we damp out any residual radial motions artificially, assuming the spun-up star is in a quasi-steady state. The total angular momentum in this star is $\sim6\times10^{54}$~g\,cm$^2$\,s$^{-1}$, comparable to the amount of angular momentum in the pre-merging binary. The equatorial extent of the bloated envelope reaches up to $\sim1000~\rsun$, which has surface escape velocities of $\sim150$~km~s$^{-1}$. This means that the stellar winds emitted during this phase could have terminal velocities of similar magnitudes, and could correspond to the velocities inferred from absorption lines in the light echoes \citep{res12,pri14,smi18b}.

The outer parts of this spun-up star resembles the outer parts of common envelopes in 3D simulations fairly well \citep[e.g.][]{ohl16,iac17,pej17,mac18,rei19,schr20}. However, it differs from a real common-envelope situation in the central region. In reality there are two cores orbiting each other within the envelope and is gradually falling in, whereas the artificially spun-up star simply has a spinning core. The spiral-in time-scale can be estimated by calculating the drag force acting on the cores. We take into account two different types of drags. One is due to ram pressure
\begin{equation}
 F_\mathrm{drag,ram}\sim \frac{1}{2}\rho v_\mathrm{rel}^2C_\mathrm{drag}A,
\end{equation}
where $\rho$ is ambient density, $A$ is the cross sectional area and $C_\mathrm{drag}$ is a drag coefficient which we take as $1/3$ which takes into account stellar compression \citep{RH18}. $v_\mathrm{rel}=v_\mathrm{orb}-v_\varphi$ is the relative velocity between the cores and the rotating envelope. Assuming that the orbit shrinks purely due to this drag force, the spiral-in time-scale becomes
\begin{equation}
 \tau_\mathrm{spiral,ram}=\frac{GM_\mathrm{1,c}M_\mathrm{2,c}}{a\rho v_\mathrm{rel}^3C_\mathrm{drag}A},
\end{equation}
where $G$ is the gravitational constant, $M_\mathrm{1,c}, M_\mathrm{2,c}$ are the masses of the cores and $a$ is the orbital separation. Normalizing this by the orbital period, it becomes
\begin{equation}
% \tau_\mathrm{spiral,ram}'\equiv \frac{\tau_\mathrm{spiral}}{P_\mathrm{orb}}=\frac{G^{3/2}M_\mathrm{cores}^{5/2}x(1-x)}{2\pi a^{5/2}\rho v_\mathrm{rel}^3C_\mathrm{drag}A},
 \bar{\tau}_\mathrm{spiral,ram}\equiv\frac{\tau_\mathrm{spiral}}{P_\mathrm{orb}}=\frac{q}{(1+q)^2}\frac{G^{3/2}M_\mathrm{cores}^{5/2}}{2\pi a^{5/2}\rho v_\mathrm{rel}^3C_\mathrm{drag}A},
\end{equation}
where $M_\mathrm{cores}\equiv M_\mathrm{1,c}+M_\mathrm{2,c}$ is the total mass of the cores and $q\equiv M_\mathrm{2,c}/M_\mathrm{1,c}$ is the mass ratio of the cores. 

We also compute the drag due to dynamical friction \citep[]{ost99}
\begin{eqnarray}
 F_\mathrm{drag,DF}=-\frac{4\pi(GM_\mathrm{2,c})^2\rho}{v_\mathrm{rel}^2}I(\mathcal{M}),
\end{eqnarray}
where $\mathcal{M}$ is the Mach number and
\begin{eqnarray}
 I(\mathcal{M})=\begin{cases}
		 \dfrac{1}{2}\ln{\left[\dfrac{1+\mathcal{M}}{1-\mathcal{M}}\right]}-\mathcal{M},& \mathcal{M}<1\\
		 \dfrac{1}{2}\ln{\left[1-\dfrac{1}{\mathcal{M}^2}\right]}+\ln{\left[\dfrac{2rv_\mathrm{rel}^2}{GM_\mathrm{cores}}\right]} ,& \mathcal{M}>1
		\end{cases}
\end{eqnarray}
where the second term in the supersonic case is taken from \citet{gin20}. This leads to a normalized spiral-in time-scale of
\begin{eqnarray}
 \bar{\tau}_\mathrm{spiral,DF}=\frac{1}{q}\frac{M_\mathrm{cores}^{1/2}v_\mathrm{rel}}{16\pi^2a^{5/2}G^{1/2}\rho I(\mathcal{M})}.
\end{eqnarray}

\begin{figure}
 \centering
 \includegraphics[]{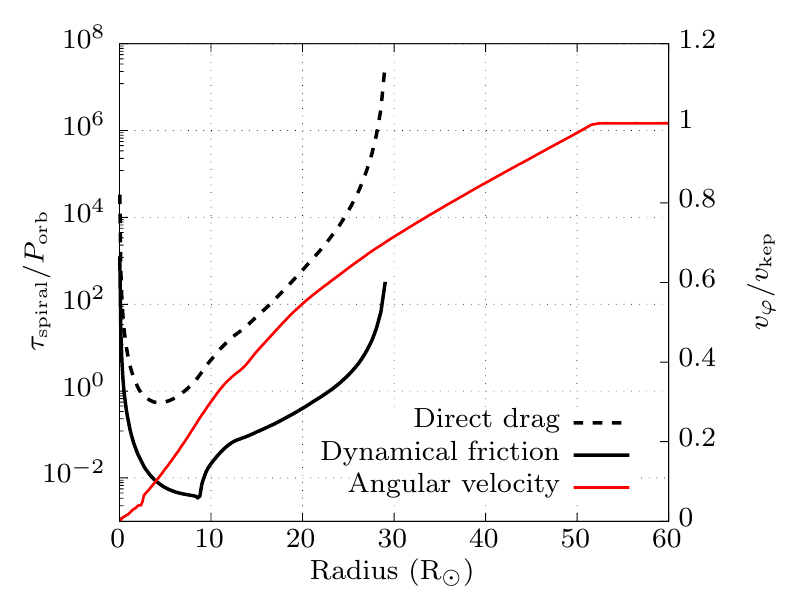}
 \caption{Spiral-in time-scale and angular velocity at each radius in the equatorial plane of the spun-up star. The solid curve is the spiral-in time-scale calculated with dynamical friction and the dashed curve is calculated with direct drag. Here we assumed $A=\pi(5~\rsun)^2$, $M_\mathrm{cores}=70~\msun$, the mass of the cores to be equal ($q=1$) and therefore the orbital velocity is $v_\mathrm{orb}=\sqrt{|\phi|/2}$ where $\phi$ is the local gravitational potential.\label{fig:spiralintime}}
\end{figure}

In Figure~\ref{fig:spiralintime} we show the spiral-in time-scales and angular velocity as a function of radius on the equatorial plane of the spun-up star assuming $q=1$. The angular velocity is normalized by the local Keplerian velocity $v_\mathrm{kep}=\sqrt{|\phi|}$. Everything outside $r\gtrsim50~\rsun$ is rotating at the local Keplerian velocity and is uniformly rotating inside due to the construction. Because the binary separation is twice the local radius from the centre of mass, the orbital velocity of the binary is a factor $\sqrt{2}$ times smaller than the local Keplerian velocity. Therefore the binary will synchronously rotate with the envelope if it is placed at a separation of $a\sim30~\rsun$ where $v_\varphi=v_\mathrm{kep}/\sqrt{2}$. If the binary separation is smaller, the orbit is faster than the local envelope rotation and thus a drag force acts on it to further shrink the orbit. Dynamical friction dominates over the direct drag at all positions. The spiral-in time-scale becomes comparable to the orbital period (i.e. $\bar{\tau}_\mathrm{spiral}\sim\mathcal{O}(1)$) when the core binary has a separation of $a\lesssim40~\rsun$ ($r\lesssim20~\rsun$). We will assume that the dynamical phase starts once the core binary has shrunk to a separation $a\leq a_\textsc{dp}$.

The spun-up star in our simulation has a centrally concentrated single core while it should have a core binary in a real common-envelope situation. So the total energy and angular momentum in the central $r<a_\textsc{dp}/2$ region are significantly lower in our simulation. We simulate the final dynamical phase by artificially filling in this gap through rapid angular momentum and energy injection. This is intended to mimic the energy and angular momentum release in the tidal disruption or violent merger of the cores. The true total angular momentum and energy budget of the system can be estimated by
\begin{align}
& J_\mathrm{tot,real}=\int_{r>a_\textsc{dp}/2}r\sin{\theta} \rho v_\varphi dV + J_\mathrm{orb},\\
& E_\mathrm{tot,real}=\int_{r>a_\textsc{dp}/2}\left(\frac{1}{2}\rho\phi+e\right)dV+E_\mathrm{orb},
\end{align}
where the integrals are taken over the envelope of the spun-up star and
\begin{align}
% J_\mathrm{orb}=x(1-x)\sqrt{\frac{GM_\mathrm{cores}^3}{a_\textsc{dp}}},\\
% J_\mathrm{orb}=\frac{q}{(1+q)^2}\sqrt{\frac{GM_\mathrm{cores}^3}{a_\textsc{dp}}},\\
 &J_\mathrm{orb}=\frac{q}{(1+q)^2}\sqrt{GM_\mathrm{cores}^3a_\textsc{dp}},\\
 &E_\mathrm{orb}=-\frac{q}{(1+q)^2}\frac{GM_\mathrm{cores}^2}{2a_\textsc{dp}},\\
 &M_\mathrm{cores}=\int_{r\leq a_\textsc{dp}/2}\rho dV.
\end{align}
For simplicity, we assume $q=1$ which gives the largest energy and angular momentum budget. We first inject angular momentum by applying
\begin{equation}
 \dot{\Omega}_\mathrm{add}=\frac{J_\mathrm{orb}}{I_\mathrm{core}T_\mathrm{inj}},
\end{equation}
to everywhere in the envelope ($r>a_\textsc{dp}/2$) that is sub-Keplerian ($v_\varphi^2<|\phi|$). Here, $I_\mathrm{core}$ is the moment of inertia of the region inside $r\leq a_\textsc{dp}/2$ and $T_\mathrm{inj}$ is an injection time-scale which we set to a fraction of the orbital period. Once the total angular momentum in the simulation reaches $J_\mathrm{tot}=J_\mathrm{tot,real}$, we switch to $\dot{\Omega}_\mathrm{add}=0$ and then impulsively add internal energy to a shell-like region. The amount of energy injection is chosen so that the total amount of energy in the computation after injection becomes $E_\mathrm{tot}=E_\mathrm{tot,real}$. This procedure makes sure that the total angular momentum and energy does not exceed the amount available in the system. Note that angular momentum injection already adds some kinetic energy to the system and the amount depends on the choice of $T_\mathrm{inj}$ because the momentum of inertia changes during the injection. Choosing longer $T_\mathrm{inj}$ leads to lower kinetic energy and therefore more of the energy will be injected as internal energy. Our fiducial injection region ($a_\textsc{dp}/2\leq r\leq a_\textsc{dp}$) assumes that most of the energy will be dissipated around the tidal disruption radius, but we also ran models with different injection regions for comparison. 

Total energy and angular momentum rises during the injection, but after switching off the injection, we checked that both the total energy and angular momentum is conserved within $\lesssim0.3\%$ in our simulations. 

The various model parameters used for the simulations are summarized in Table \ref{tab:eruption_results} along with part of the results. $\Delta J$ shows how much angular momentum was injected in the dynamical phase. $\Delta E_\mathrm{kin}$ and $\Delta E_\mathrm{int}$ show the amount of energy injected in the form of kinetic energy and internal energy respectively. The sum of the energies are the same for models with the same $a_\textsc{dp}$ parameter. For the models with uniform energy injection per unit volume, most of the injected internal energy will be located at the outer edge of the injection radius (higher mass coordinate) whereas for models with uniform energy injection per unit mass, the majority of the energy is injected at the inner edge (lower mass coordinate). Also, the choice of $T_\mathrm{inj}$ affects the effective mass coordinate of the energy injection too because the injection radius is fixed in space, so mass can flow out of the injection region during the angular momentum injection phase.

\begin{table*}
 \caption{Model parameters and results of the eruption simulations.}
 \begin{center}
  \begin{tabular}{ccccccccc}
   \hline
   Model & $a_\textsc{dp}$ & Injection radius & $T_\mathrm{inj}$ & $\Delta J$ & $\Delta E_\mathrm{kin}$ & $\Delta E_\mathrm{int}$ & Ejecta mass & Ejecta energy \\
   & ($\rsun$) & ($\rsun$) & (ks) & (g~cm$^2$~s$^{-1}$) & (erg) & (erg) & ($\msun$) & (erg) \\\hline
   \noalign{\vspace{2pt}}
   1 & 20 &  10--20$^a$ & 14 & $3.02\times10^{54}$ & $2.13\times10^{50}$ & $7.71\times10^{49}$ & 16.9 & $9.2\times10^{49}$ \\
   2 & 20 &  10--20$^a$ & 28 & $3.02\times10^{54}$ & $1.63\times10^{50}$ & $1.27\times10^{50}$ & 16.1 & $8.7\times10^{49}$ \\
   3 & 15 & 7.5--15$^a$ &  5 & $1.82\times10^{54}$ & $1.09\times10^{50}$ & $1.02\times10^{50}$ &  8.7 & $3.6\times10^{49}$ \\
   4 & 15 & 7.5--15$^a$ & 13 & $1.82\times10^{54}$ & $1.02\times10^{50}$ & $1.09\times10^{50}$ &  8.1 & $3.0\times10^{49}$ \\
   5 & 20 &   0--10$^a$ & 14 & $3.02\times10^{54}$ & $2.13\times10^{49}$ & $7.71\times10^{50}$ & 11.8 & $4.7\times10^{49}$ \\
   6 & 20 &  10--20$^b$ & 28 & $3.02\times10^{54}$ & $1.63\times10^{50}$ & $1.27\times10^{50}$ & 15.8 & $5.9\times10^{49}$ \\
   7 & 20 &   5--15$^a$ & 28 & $3.02\times10^{54}$ & $1.63\times10^{50}$ & $1.27\times10^{50}$ & 13.1 & $4.6\times10^{49}$ \\\hline
  \end{tabular}\label{tab:eruption_results}
 \end{center}
 \raggedright $^a$Uniform energy injection per unit volume.\\
              $^b$Uniform energy injection per unit mass.
\end{table*}

\begin{figure*}
 \centering
 \includegraphics[width=\linewidth]{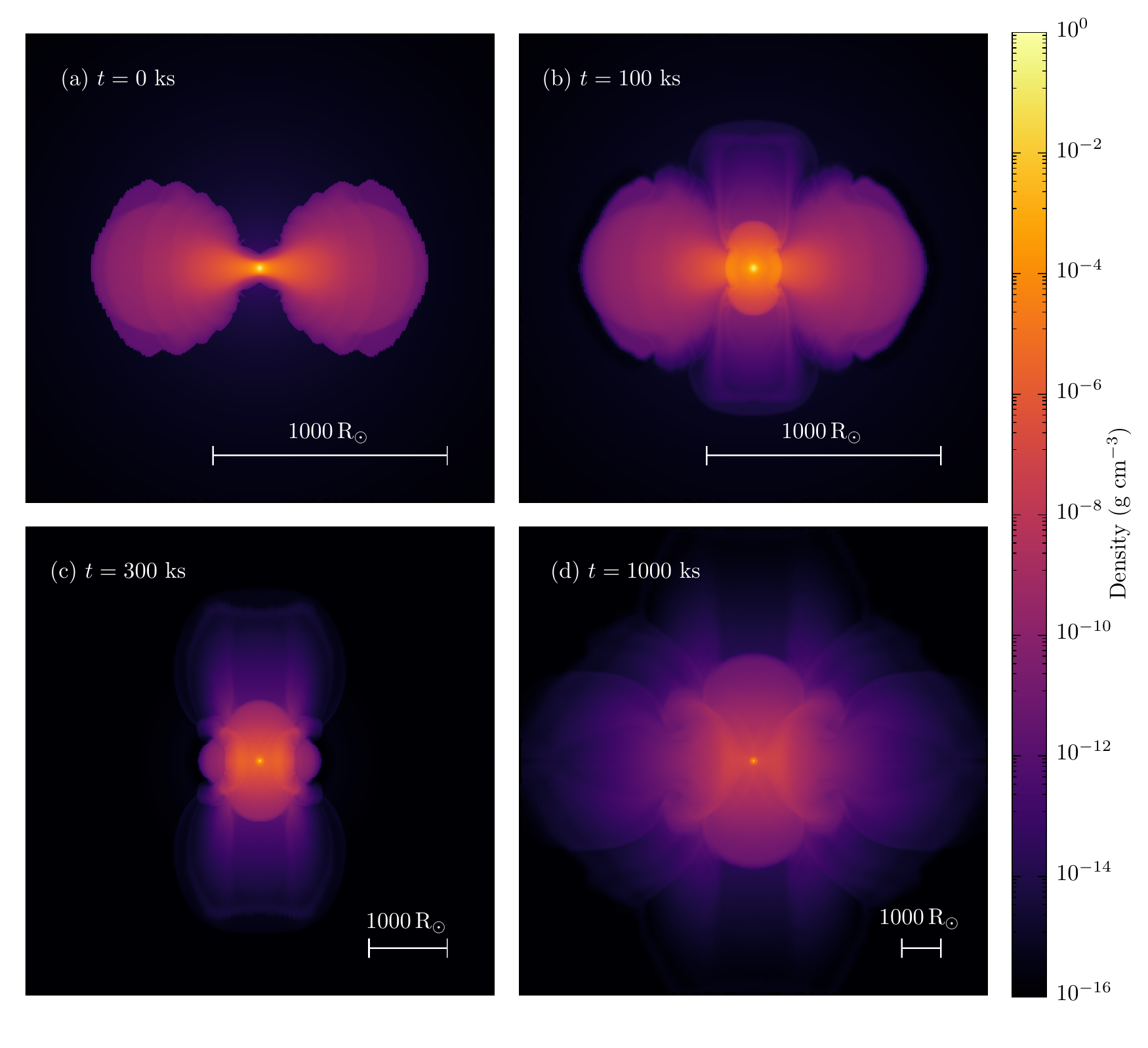}
 \caption{Density distribution snapshots of the eruption simulation for our fiducial model (Model 6). Time is counted from the beginning of the dynamical phase. Each box has a different scale. A movie of the eruption simulation is available online (Movie~1).\label{fig:snapshots_eruption}}
\end{figure*}

Figure~\ref{fig:snapshots_eruption} shows snapshots of the eruption simulation in our fiducial model (Model 6). Panel (a) shows the highly oblate spun-up star that is used as the initial condition for all eruption simulations. The injection of angular momentum and energy ends at about $\sim28$~ks and the excess energy quickly drives an outgoing shock. The shock first breaks out through the poles as in panel (b), while the equatorial part of the shock slowly propagates through the oblate envelope which can be seen in panel (c). The equatorial shock eventually reaches the surface of the torus too and breaks out at very high velocities reaching $\gtrsim10,000$~km~s$^{-1}$. After equatorial shock breakout, the ejecta simply follow a homologous expansion, keeping the relative mass distribution in panel (d). The amount of mass and energy injected in the eruption are displayed in Table \ref{tab:eruption_results}. As expected, the higher energy injection models show greater ejecta mass and energy. Longer energy injection times lead to less ejecta because in our energy injection procedure, longer injection times allow the matter to flow out of the injection sphere and therefore the energy is injected into an effectively deeper layer. Deeper injection is known to show less ejection \citep[]{owo19}. Injecting energy proportional to mass (Model 6) also leads to effectively deeper injection, so it has less ejecta mass and energy.

\begin{figure}
 \centering
 \includegraphics[]{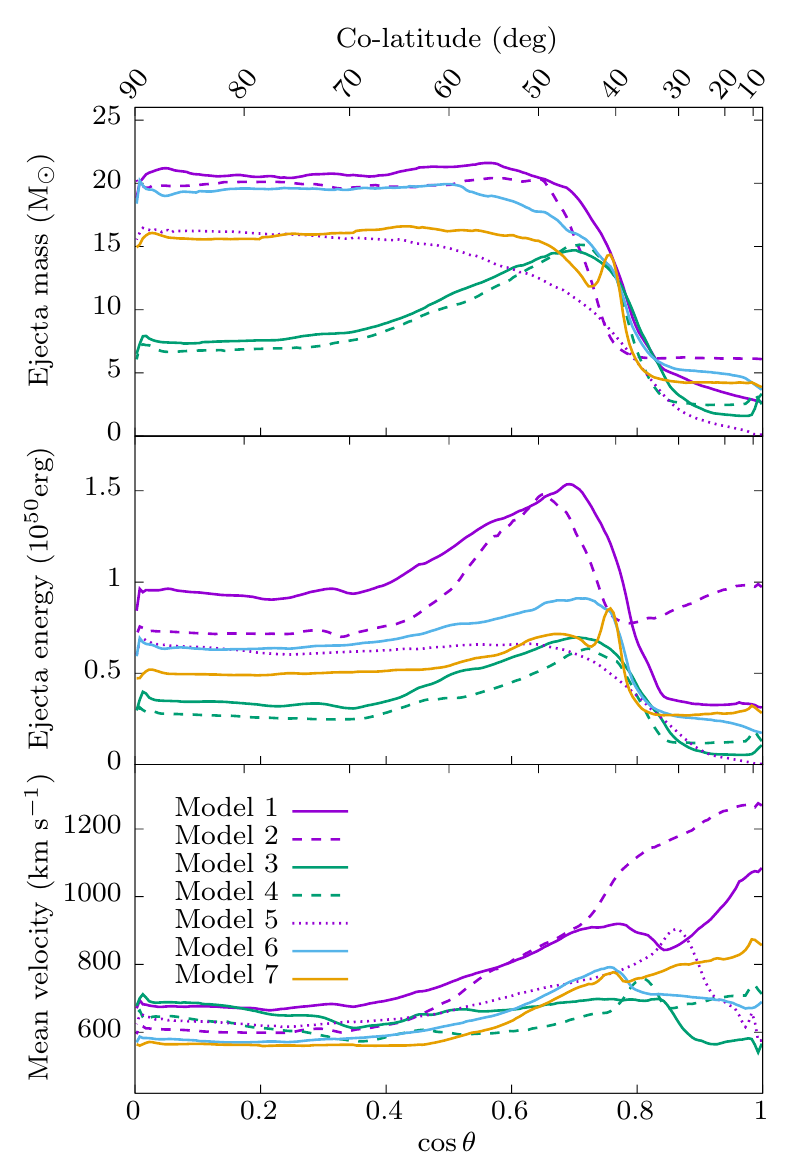}
 \caption{Latitudinal distribution of ejecta mass, energy and mean velocity. Mass and energy are normalized by the solid angle so it shows the spherical equivalent values. Mean velocity is the energy weighted mean.\label{fig:angulardist}}
\end{figure}

In Figure~\ref{fig:angulardist} we show the latitudinal distribution of ejecta mass, energy and mean velocity. The mass distribution is roughly flat around the equator and the poles, with higher mass in the equatorial region. Lower energy models have less equatorial ejecta, which is qualitatively similar to the findings in \citet{mor06}. Models with larger $T_\mathrm{inj}$ show more polar mass ejection (dashed vs solid curves). The ejecta energy shows a similar shape but peaks at around $\cos{\theta}\sim0.7$ ($\sim45\degree$), because as the shock propagates outwards, it gets deflected towards the poles by the high density around the equator. This is a feature consistent with \citet{mor06}. The mean velocity distribution is somewhat more scattered but most models have a higher velocity around the poles. This means that the ejecta are roughly bipolar, but there is less mass ejected towards the poles. Here we conclude that the eruption alone is not able to create a bipolar Homunculus but there is plenty of mass, energy and momentum available in the ejecta to explain the kinematics of the shell if it is appropriately redistributed.

The key assumption of this phase is that both of the merging stars have established a dense core-like region. This is the reason that we require the three stars in the triple system to have similar initial masses. Rapid energy dissipation upon core merger and the associated explosion-like eruption is naturally expected as long as this assumption holds \citep[e.g.][]{mor06,ro17,schr20}. Technically, the late main sequence stars we assume in this scenario do not have well-defined cores, but they do have convective cores that have higher mean molecular weight and lower entropy compared to the rest of the star. Therefore we expect that once the stars start merging, these heavy low-entropy regions will decouple from the envelope and start spiralling in inside the common envelope. Such core--envelope structures have large dynamic ranges and makes it difficult to perform full 3D simulations. This is the key difference from previous stellar merger studies \citep[e.g.][]{sch19} and is the reason why we chose a 2.5D approach.

\subsection{Sweep-up simulation}\label{sec:sweepup}

After the eruption, the star contains a huge amount of excess energy that still has to be emitted to regain thermal equilibrium. The resulting large outward energy flux can cause instabilities that transform the atmosphere into a porous medium \citep[]{sha99,beg01}. This results in a reduced effective opacity, allowing for sustained super-Eddington luminosities that can drive a strong continuum-driven wind \citep[e.g.][]{sha00,owo04}. The mass-loss rate can reach values reaching up to the maximum allowed limit (so-called ``photon tiring limit'') of $\sim 0.1 M_\odot$/yr, or about $\sim$100 times higher than what is inferred for the current-day wind \citep[]{owo17}. Moreover, the angular momentum from the merger causes the combined star to have rapid, near-critical rotation. The lower effective gravity and associated gravity darkening at lower latitudes \citep[]{zei24} leads to a wind that is slower and weaker from the equator, and faster and stronger from the poles \citep[]{cra95,owo96}. This wind will interact with the inner parts of the ejecta, sweeping it up into the thin, hollow bipolar shells we observe today as the Homunculus nebula. Here we further extend the hydrodynamical simulations to investigate how the ejecta will be swept up by a post-eruption wind. This is somewhat similar to the ``snow plow'' model for pulsar wind nebulae \citep[]{ost71,che92}, except that the ejecta and wind are assumed to be much more aspherical for our case. Similar attempts, simulating the interactions between different wind phases, have been made in the past \citep[]{fra95,fra98,gar97,dwa98,lan99,gon04a,gon04b,gon10,gon18}. Real explosions like we have simulated above do not have constant velocity distributions like steady winds but have linear distributions. The density distribution in explosion ejecta are also very different from the $\rho\propto r^{-2}$ distribution in winds \citep[]{owo19}. Our approach more closely represents the merger situation and the results are expected to be qualitatively different from the previous studies.

At the endpoint of our eruption simulation, the outer cells have positive total energy, representing the unbound ejecta whereas inner cells have negative total energy, representing the bound star. Proper modelling of the porous atmosphere and driving of the super-Eddington wind requires very expensive 3D radiation-hydrodynamic simulations. Due to the high computational demand, we postpone such attempts to future work and take a simpler empirical approach to model only the interaction between the wind and ejecta. We cut out the inner bound region from the grid and replace it with a wind blowing into the box as an inner boundary condition. At this stage the optical depth of the ejecta material has become low enough for radiation to decouple from the gas, so we drop the effect of radiation pressure in the equation of state. We also add a cooling term in the energy equation to account for radiative cooling, which should become efficient when the optical depth is low (see Appendix \ref{app:code} for details). Cooling is required to produce the thin walls of the Homunculus \citep[]{wea77,smi13}. We carry out the simulation for our fiducial eruption model (Model 6) and run it for 180~yr to compare it with what we observe today.

The injection wind strength is chosen so that the shell velocity reaches roughly $\sim650$~km~s$^{-1}$ at the pole by the end of the simulation (see Appendix \ref{app:shell} for details). For our fiducial model the wind parameters are $\dot{M}=0.1~\msun$~yr$^{-1}$ and $v_w=1000$~km~s$^{-1}$. We assume the wind scales as
\begin{align}
& \dot{M}(\theta,t)=\dot{M}_0\left(1-W_\mathrm{rot}^2\sin^2{\theta}\right)\exp\left[-\left(\frac{t}{\tau_\mathrm{dec}}\right)^2\right],\\
& v_w(\theta)=v_{w,0}\sqrt{1-W_\mathrm{rot}^2\sin^2{\theta}},
\end{align}
where $\dot{M}_0$ and $v_{w,0}$ are the initial mass-loss rate and velocity at the pole, respectively. $W_\mathrm{rot}$ is the rotational velocity normalized by the critical spin velocity, which we set to $W_\mathrm{rot}^2=$0.95. The bipolar form we assume here for velocity follows that inferred for the present-day wind, with wind speed that varies from $\sim 1000~$km~s$^{-1}$ to $\sim 600~$km~s$^{-1}$ from pole to equator, following scalings for a stellar envelope spun up by the merger to near-critical rotation \citep{cra95,owo96,owo04}. The associated equatorial gravity darkening also leads to a polar-enhanced mass flux, which is likewise inferred in the present-day wind \citep{smi03a}. We also assume that the mass-loss rate decayed over a time-scale of $\tau_\mathrm{dec}\sim90$~yr so that it matches the current mass-loss rate ($\dot{M}\sim10^{-3}~\msun$~yr$^{-1}$). A decay in the mass-loss rate is in fact observed \citep[]{meh10,mad13}.

\begin{figure}
 \centering
 \includegraphics[]{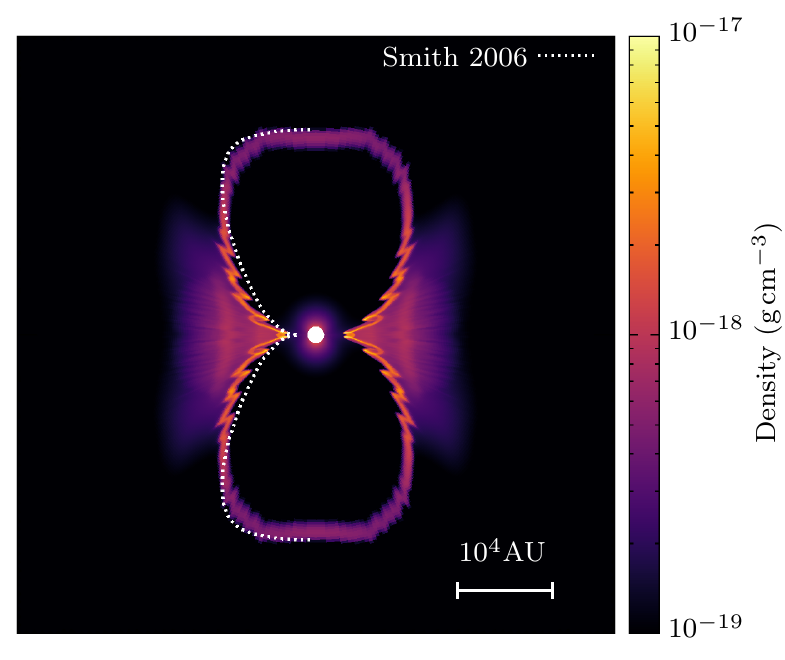}
 \caption{Density snapshot of the sweep-up simulation at 160~yr after eruption. The white dotted curve indicates the observed Homunculus shape taken from \citet{smi06a}. A movie of the sweep-up simulation is available online (Movie~2).\label{fig:finalshell}}
\end{figure}

In Figure~\ref{fig:finalshell} we show the final snapshot of our fiducial model. There is a clear bipolar thin shell that is very similar to what is seen in \etac. The white dotted curve shows the observed Homunculus shape determined by \citet{smi06a} for comparison. The overall shape of the simulated shell closely resembles that of the observed Homunculus. There are some jagged features along the shell, which is likely due to the ``thin-shell instability'' or ``Vishniac instability'' \citep[]{vis83,kee14}. However, such features may be due to the strict enforcement of axisymmetry in our simulation and could be smeared out by azimuthal motions in a real 3D case, or more realistic treatments for cooling \citep[]{bad16}. It also may reproduce the observed mottling of the shells \citep[]{smi13}. Note that the jagged spikes are each resolved with >10 polar grid points and are not reflecting grid-size effects.

\begin{figure}
 \centering
 \includegraphics[]{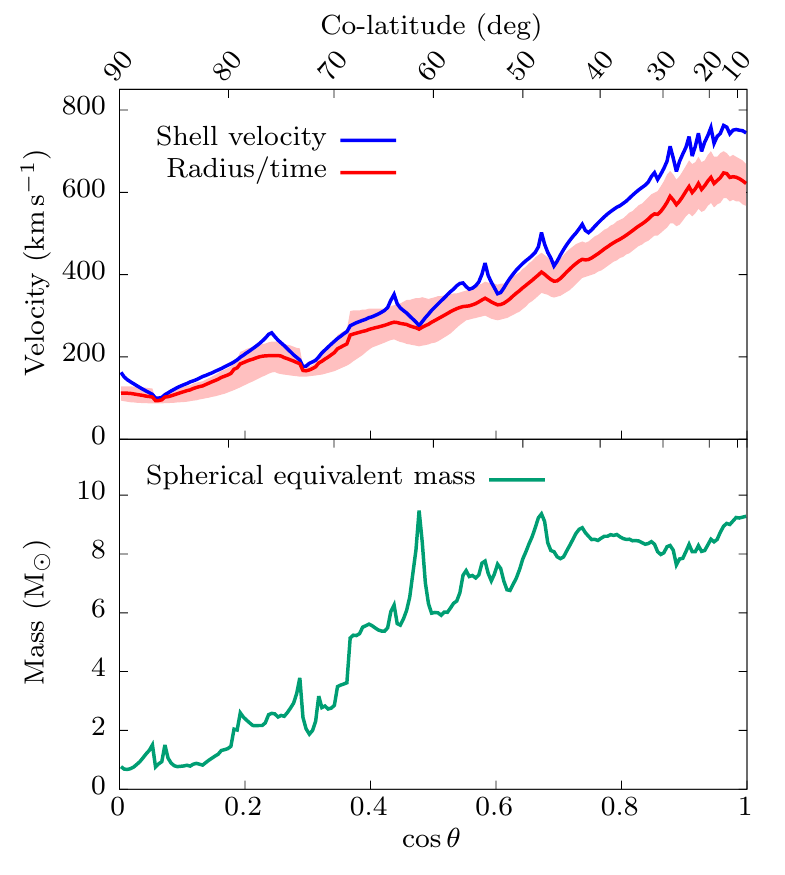}
 \caption{Latitudinal distribution of position, velocity and mass of the shell. The radius is the mass-averaged shell radius and is divided by the time after eruption ($t=160$~yr) to give a dimension of velocity. The pink shaded region shows the width of the shell.\label{fig:shellstats}}
\end{figure}

To analyse our results more quantitatively, we use the following strategy to identify the shell. The velocity distribution along each radial direction can be divided into three parts; the wind, the shell and the ejecta. The wind has a constant velocity and the ejecta has a linear distribution ($v=r/t$), and there is a transitional region in between. We define this transitional region as the shell and calculate the mass-weighted radius and velocity along each latitudinal ray. The total mass in the shell is $\sim5.5~\msun$, which is smaller than the mass estimates from observations \citep[$\gtrsim10~\msun$;][]{smi06a}.

The latitudinal distributions are shown in Figure~\ref{fig:shellstats}. One remarkable feature is that the shell velocity is 10--20$\%$ larger than its position divided by time. This means that the shell does not follow a clean homologous expansion and there is a velocity difference between the shell and the ejecta material right above it\footnote{This corresponds to $\eta<1$ in the analysis presented in Appendix \ref{app:shell}.}. Such a difference will alter the apparent ejection date of the shell to slightly (a few decades) after the Great Eruption. The ejection dates inferred from proper motions points to 1847$\pm$1, which is a few years after the peak of the Great Eruption \citep[]{smi17}. Although the discrepancy is much smaller, this is qualitatively in agreement with our model. Any scenario that involves deceleration of Great Eruption ejecta by pre-eruption material will have an apparent ejection date ``before'' the Great Eruption. The velocity differences decrease towards the equator, implying a latitude-dependent age discrepancy. The width of the shell is 10--15$\%$ of the shell radius, but this strongly depends on the treatment of cooling, hence we do not claim this is a robust quantity.

The mass distribution within the shell is also interesting. Although in Figure~\ref{fig:finalshell} the density visually looks higher at lower latitudes, there is more mass around the poles when integrated over the width of the shell. The mass distribution is almost constant up to $\sim50\degree$ and drops off towards the equator. This is opposite from the mass distribution of the ejecta where there was more mass towards the equator. It also seems to be consistent with the observations that the polar caps are more opaque compared to the side walls \citep[e.g.][]{hil92,dav01,smith02}. However, the mass distribution within the Homunculus walls is sensitive to the details of the eruption and post-eruption winds, and the distribution we show here is not a robust result. The opacity distribution also strongly depends on the dust formation process and we do not attempt to model that in this paper.

There is also a lot of ejecta material ($\sim10~\msun$) outside the Homunculus in the low-latitude, low-velocity regions that is not yet swept up by the shell.  Ultra-violet observations of resonant scattering from Mg~\textsc{ii} also show matter outside the Homunculus that spatially coincides with the extended material in our simulations \citep{smi19}, although the current estimated mass ($\gtrsim0.02~\msun$) is significantly smaller than what we obtain. The main reason for the mass enhancement in the lower latitudes is due to the oblateness of the pre-eruption envelope shaped by the large amount of angular momentum brought in from the orbit. This is a distinct feature of the merger hypothesis, and is important for the shaping of the Homunculus. However, the ratio between the Homunculus and outer masses can be somewhat tuned by changing the energy injection procedure. For example, weaker energy injection models have less ejecta in the low-latitude regions (Models 3 \& 4 in Figure~\ref{fig:angulardist}). This can still be sculpted into the Homunculus shape by invoking stronger post-eruption winds, and there will be less mass outside the Homunculus since the lower velocity ejecta will be swept up into the shell. Indeed, there are some observational indications that there may be a mass of $5~\msun$ or more that is already swept up in the equatorial waist of the Homunculus nebula \citep{mor17,smi18c}, although with large uncertainties. This could have originally been the lower velocity eruption material that has been swept up by the wind and contributed to the pinching of the Homunculus equator. Therefore, further detailed observational and theoretical investigations of the mass and its distribution of the matter outside the Homunculus may help us understand the structure of the pre-merger envelope, the true nature of the energy deposition and the relative contributions of the eruption and wind for shaping the Homunculus nebula.

In this scenario, the shell should still be slightly accelerating depending on what the wind strength was $\sim110$~yr ago. For example in the simulation we show here, the Homunculus expansion speed accelerates by $\sim1~\%$ from 2005 to 2020.   This could in principle be verified by future detailed observations of the Homunculus.

\section{Historical ejections}\label{sec:sprays}
The Outer Ejecta showing pre-eruption mass loss in multiple precursor eruptions over several hundred years \citep{kim16} has been difficult to reconcile with a simple binary merger scenario. A terminal event like a binary merger will in general not produce recursive mass-loss events and therefore requires other mechanisms for pre-eruption mass loss. \citet{smi18b} proposed that in a triple system, $\eta$~Car's precursor eruptions that made the Outer Ejecta arose from the interaction of two of the massive stars grazing each others surfaces at high eccentricities, based on studies of unstable orbital dynamics in triples \citep[]{per12}. The eccentricity can be periodically excited through triple body interactions \citep[]{per12,sha13,mic14}. In this section we explore the possibility of mass ejections in eccentric orbits and the expected distribution of these ejectiles around the system. We then compare it with various observational properties of the historical ejecta.

There are several observational properties of the outer ejectiles that need to be addressed in any theoretical model for \etac's formation. Based on the ejection dates inferred from the proper motion, there are at least 3 distinct ejection episodes at around the years 1250, 1550 and 1800 A.D. \citep[]{kim16}. The velocity range is 300--600~km~s$^{-1}$\footnote{Note that these are projected velocities so the physical velocities should be larger.}. Each ejection episode has a different ejection direction which is not aligned to the Homunculus symmetry axis and, more importantly, is not even bipolar. It rather looks like one-sided sprays of ejecta with some opening angle. Hereafter we will call the historical Outer Ejecta as the ``sprays''. Any model for the spray ejection has to self-consistently explain the $\sim$300~yr intervals, the 300--600~km~s$^{-1}$ velocities and the disorganised ejection directions.

\subsection{Orbital evolution towards merger}\label{sec:triple}
Because our scenario assumes that the Homunculus was created through a binary merger, the existence of a current-day companion naturally requires the system to have been a triple prior to the Great Eruption. The possibility of {\etac} being a triple has been raised in the past \citep[]{liv98}, and some studies suggest that the triple interactions can cause very close encounters of the stellar components and ultimately trigger a coalescence \citep[]{por16,smi18b}. However, the current day orientation of the binary orbit to the Homunculus nebula complicates the problem. The current day orbit of the companion appears to be well aligned with the Homunculus equatorial plane \citep[]{gul09,mad12}. If the Homunculus symmetry axis was determined by the orbit of the merging stars, it would mean that the pre-merger triple system had a small mutual inclination. In normal Kozai-Lidov oscillations the eccentricity reaches high values for systems with large mutual inclination \citep[]{koz62,lid62}, so it is expected that the mutual inclination at the time of merger is large. In fact, The Kozai-Lidov mechanism only works for systems with mutual inclinations of $i>\arccos{\sqrt{3/5}}\sim39.2\degree$; i.e. the inclination between the merger plane and outer orbit cannot be smaller than this value.\footnote{Indeed, \citet{por16} use mutual inclinations of $i\sim90\degree$.}. There are some processes where the mutual inclination can be damped after the merger, such as tidal interactions of the inner and outer orbits in the spiral-in phase \citep[]{cor13} or partial alignment of the spun-up envelope through disk-orbit interaction \citep[]{mar11}. But the time-scales of tidal processes tend to be quite long, so it is unlikely that they reach the observed low inclinations ($\sim0\degree$).

Another puzzling factor is the nature of the companion star itself. The hard X-ray emission at periastron suggests strong colliding wind interactions, implying that the companion wind should have a comparable momentum to that of {\etac} \citep[]{cor95,cor97}. Hydrodynamical modelling estimates the wind properties of the present-day companion to be $\dot{M}\sim10^{-5}~\msun$~yr$^{-1}$ and $v_\infty\sim3000$~km~s$^{-1}$ \citep[]{oka08,rus16,ham18}, which is orders of magnitude stronger and several times faster than the typical wind of a $\sim30~\msun$ main-sequence star. In fact, it is more consistent with a hydrogen-poor star like Wolf-Rayet (WR) stars \citep{smi18b}. 

This issue has been discussed by \citet{smi18b}, and they construct a speculative model that could possibly resolve this. In their model the current companion to {\etac} was initially the most massive star in the triple system which is eventually kicked out to be the tertiary star after losing its hydrogen envelope through mass transfer. In this way the WR-like nature of the current companion and eccentric orbit can naturally be explained. 

To investigate this model we consider the evolution of a hierarchical triple system with masses of $M_A\sim60~\msun, M_B\sim40~\msun$ and $M_C\sim50~\msun$ for stars A, B and C respectively. We assume that the mutual inclination is initially large ($>40\degree$). The system is thus subject to Kozai-Lidov oscillations but the distances at periastron do not get close enough to cause catastrophic interactions. Once star A reaches the end of core hydrogen burning, the star rapidly expands and starts transferring matter to star B. There is a rapid mass-transfer phase until the mass ratio inverts where the donor becomes less massive than the accretor. Part of the mass transferred in this initial short phase may be spilled out from the system because the secondary cannot accrete the transferred matter fast enough. As the mass ratio inverts, the mass transfer becomes stable and thus most of the transferred mass thereafter is expected to be accreted by the secondary. In this situation, mass transfer widens the separation of the inner binary due to angular-momentum conservation.  \citet{smi18b} proposed that this widening of the orbit after mass transfer leads to chaotic orbital evolution and an exchange of partners. Eventually star A loses most of its hydrogen envelope and so develops a strong, fast wind associated with hydrogen-depleted WR stars. This extra mass-loss stage may widen the separation slightly more, further destabilising the system. Unstable systems can undergo chaotic evolution, resulting in various outcomes such as swapping companions, ejecting stars, complete dissociation, or mergers \citep[e.g.][]{egg96,per09,per12,ant12,ant14,ant16}. Since star A is hydrogen poor at this stage, the radius is significantly smaller than that of the other two stars. Therefore the larger cross section of the other two stars make them more likely to merge with each other. The hydrogen-poor tertiary star then becomes the present-day secondary star, which may naturally explain the observed strong WR-like wind of the companion \citep[]{smi18b}.

 In chaotic situations like this, the closest encounters are not necessarily correlated with the mutual inclination, since the orbits are not well defined. Although it is tempting to give a natural explanation for the alignment of the current day orbit with the Homunculus nebula, here we will just note that in chaotic encounters it is possible to create systems with low mutual inclinations that cannot be achieved in the standard Kozai-Lidov mechanisms.

We carry out 3-body dynamical simulations to study the above scenario. For this we use a few-body integrator that directly integrates the gravitational forces with an 8th-order implicit Runge-Kutta method \citep[]{hai00}, using the coefficients of Kuntzmann \& Butcher \citep[]{but64}. With this code we follow the dynamical evolution of a marginally unstable triple system. A triple system becomes dynamically unstable when $Q\equiv a_\mathrm{out}(1-e_\mathrm{out})/a_\mathrm{in}<Q_\mathrm{st}$ where $a_\mathrm{in}, a_\mathrm{out}$ are the orbital separations of the inner and outer orbits respectively and $e_\mathrm{out}$ is the eccentricity of the outer orbit. $Q_\mathrm{st}$ is a threshold value that can be estimated by 
\begin{align}
 Q_\mathrm{st}=3\left(1+\frac{M_C}{M_A+M_B}\right)^\frac{1}{3}&\left(1-e_\mathrm{out}\right)^{-\frac{1}{6}}\nonumber\\
\times&\left(\frac{7}{4}+\frac{1}{2}\cos{i}-\cos^2{i}\right)^\frac{1}{3},\label{eq:stability}
\end{align}
based on fits to numerical experiments \citep[][see also \citealt{mar08} for analytic discussions]{val08}. This fit serves as an upper limit for instability to occur, so most systems become unstable at slightly smaller values of $Q$ \citep[see Figures 7--11 in][]{val08}. We set the initial condition of our triple system assuming that the system recently reached low $Q$ values due to mass transfer/loss. 

\begin{figure}
 \centering
 \includegraphics[]{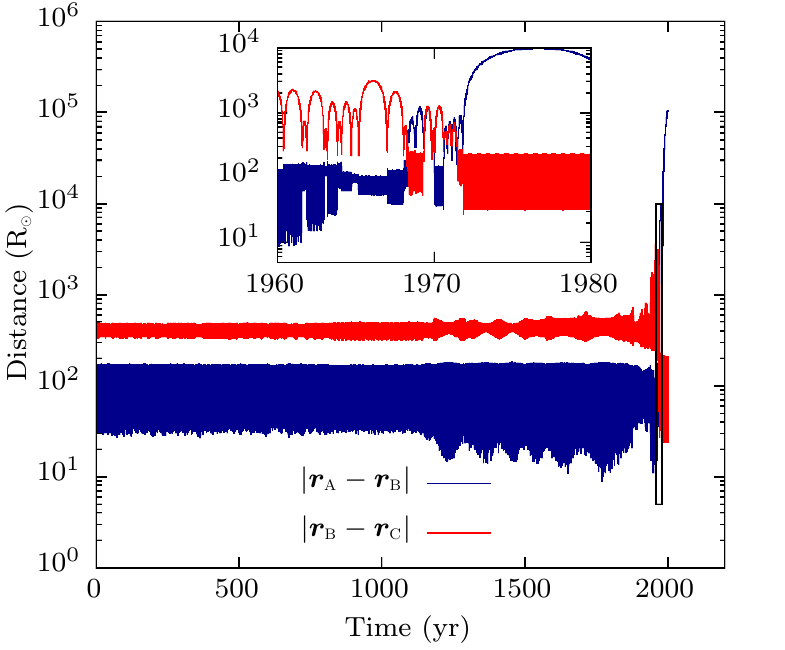}
 \caption{Time evolution of the distance between stars in an unstable triple system. The blue curve plots the distance between stars A and B, whereas the red curve plots the distance between stars B and C. Initial parameters for the simulation are $M_A=30~\msun, M_B=60~\msun, M_C=50~\msun, e_\mathrm{in}=0, a_\mathrm{in}=100~\rsun, e_\mathrm{out}=0.1, a_\mathrm{out}=400~\rsun, i=60\degree$, which correspond to the initial parameters of Phase 2-2 in Figure~\ref{fig:framework}. The inset is a zoomed-in view of the region indicated by the black box.\label{fig:distance_timeevo}}
\end{figure}

\begin{figure}
 \centering
 \includegraphics[width=3.15in]{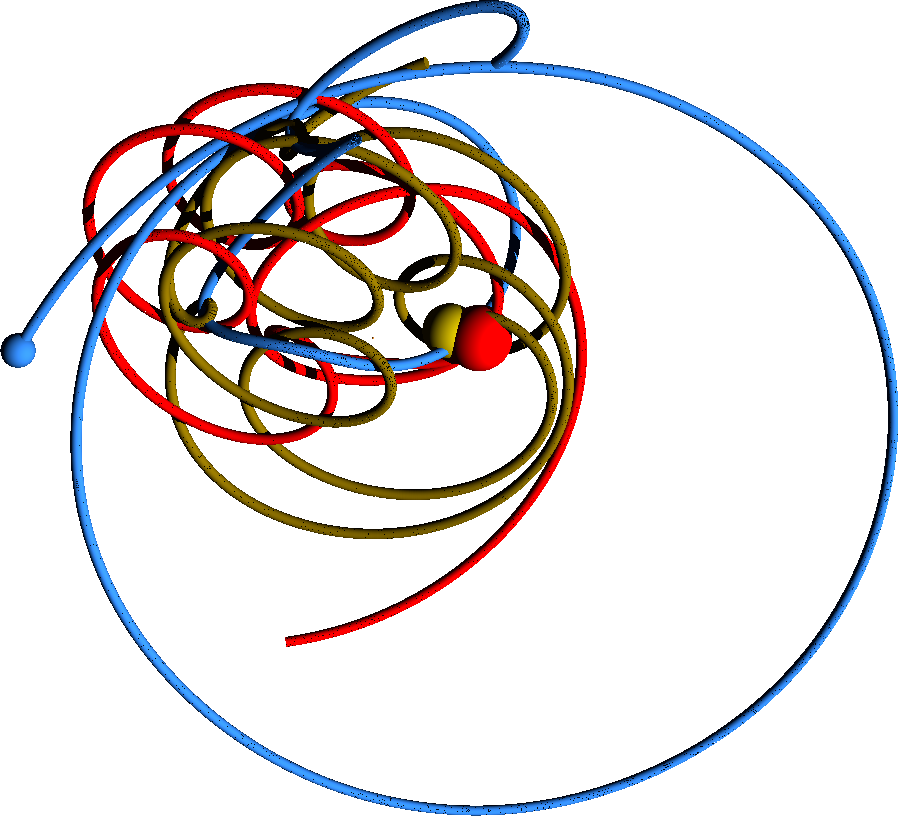}
 \caption{Trajectories of stars in the last 6 months leading up to merger. Blue, red and yellow trajectories correspond to stars A, B and C, respectively. Spheres mark the location of the stars at the point of merger. A movie of the trajectories is available online (Movie~3).\label{fig:3D_trajectories}}
\end{figure}

Figure~\ref{fig:distance_timeevo} displays an example of the time evolution of distances between bodies in an unstable triple system. The initial parameters are given in the caption, which corresponds to $Q/Q_\mathrm{st}=0.85$. The system is relatively stable for the first $\sim1000$~yr, where the inner eccentricity oscillates between 0 and 0.7 due to the Kozai-Lidov mechanism. After $\sim1100$~yr, the quasi-secular regime commences and causes modulations in the minimum distances (maximum eccentricities) reached in the Kozai-Lidov cycles. It is not clear what sets the time-scales of these modulations, but it should be pointed out that the intervals between the minimum distance peaks are $\sim100$~yr for this case, not far off from the $\sim300$~yr intervals of the observed sprays of \etac. The minimum distances reached in this chaotic phase is $\sim13$--$15~\rsun$ ($e_\mathrm{in}\sim0.85$--0.89), which is comparable to the stellar radius of a $60~\msun$ main-sequence star. Also, the minimum distance peaks are slightly deeper at later times. It may be possible that star A plunges into the envelope of star B several times, but the density in the radiative envelopes of massive stars are extremely low and therefore the drag force that acts on a WR star companion is minute unless it plunges even deeper, such as down to the surface of the convective core (See Appendix~\ref{app:drag} for discussions on how deep the WR needs to plunge in). At about $\sim1900$~yr, the system becomes even more chaotic and multiple swaps of companions occur at around $\sim1970$~yr. The minimum distance between stars B and C reach as low as $\sim20~\rsun$. Unlike star A, star C has a large radiative envelope, so the interaction with star B's envelope can be more violent than interacting with a small hydrogen-poor star. Therefore the minimum distance in this case may be small enough to trigger a catastrophic merger. The time-scale of the system reaching this disruptive point is of comparable scale to analytic estimates derived from simple random walk models \citep[$\sim4000$~yr;][]{mus20}. Figure~\ref{fig:3D_trajectories} illustrates the complexity of the orbits in the last 6 months leading to the onset of merger. If we assume that stars B and C started merging at this point ($\sim1972$~yr), the inclination between the Homunculus nebula and the current day orbit will be $i=20\degree$ or slightly smaller due to post-merger damping effects. Considering that the 3D orientation of the current day orbit still contains $\sim15\degree$ uncertainties \citep[]{mad12}, this small mutual inclination may already be small enough. The post-merger eccentricity is $\sim0.94$, which is also consistent with current estimates of \etac's orbit \citep[e.g.][]{kas07}.

As a plausibility check, we conducted the same 3-body calculation but with $a_\mathrm{in}=90~\rsun$ instead of $100~\rsun$. This corresponds to a slightly higher $Q$ value ($Q\sim0.94~Q_\mathrm{st}$), and found that this system was stable for $10^4$~yr. This means that the system was stable before the widening phase so that the chaotic interactions do not occur in the earlier stages of its evolution. The instability only started after star A evolved off the main sequence, initiating mass transfer and widening the inner orbit. This is also consistent with the 3--4~Myr age of the Tr~16 cluster\footnote{Recent parallax measurements suggest that {\etac} is unrelated to Tr~16 \citep[]{dav18}.} \citep[]{wal12}.

These numerical simulations confirm that the model proposed by \citet{smi18b} is quantitatively plausible. It remains to be determined, though, how common or rare such a scenario may be. The results we present here are only for one particular set of initial conditions as a proof of concept. Small deviations in the initial conditions (e.g. masses, inclination, orbital phase) can result in qualitatively different outcomes due to the chaotic nature of the system. We have carried out 1000 simulations of the same triple system but with variations in the initial orbital phase of the inner binary. Out of our runs, $\sim$26\% of them resulted in a stellar merger within 3000~yr. Only 2--3\% experienced companion swaps within the same time frame. A wider statistical analysis is required to quantify whether the modulating Kozai-Lidov cycles, swapping of companion and a final merger is a preferred outcome or a rare case. We will postpone such analyses to future work.

\subsection{Mass ejection from close periastron encounters}\label{sec:spray_sim}

When the eccentricity reaches extremely high values, the periastron distance can become comparable to the stellar radius and can sometimes even graze the surface of each star. These close interactions will pull off and eject material in confined directions \citep[]{smi18b}. In more extreme cases, stars can plunge deep into the envelope where the frictional force can rapidly take away orbital energy and trigger a fatal merger \citep[e.g.][]{per12}.

 To explore the outcome of grazing encounters, we conduct a simple experimental calculation of how surface material on a star reacts to the gravitational force of an approaching secondary star in an eccentric orbit. For this we use the same few-body integrator as above. To represent surface material of a star, we introduce an artificial ``glue'' force around the primary star that prevents particles from falling to the centre but instead keeps it rotating around the star at a constant spin velocity at a constant radius. This artificial force is smoothly switched off once the particle is lifted off the surface of the star. We randomly place test particles on the surface of the primary star and let it rotate at either periastron orbital angular velocity or critical rotation, whichever is smaller. It is assumed that the star has been tidally spun up through multiple encounters, or is rapidly rotating due to the preceding mass accretion. 50 particles are placed on the star and we follow its orbital evolution through 500 orbits of the secondary. Whenever a test particle reaches a distance of more than 10 times the semimajor axis away from the centre of mass, we record the direction and velocity of the ejection and put it back on a random position of the surface of the star. This procedure is equivalent to calculating a single encounter with more surface particles, but maximises the efficiency of the few-body integrator. We also remove particles whenever it gets closer than $<10~\rsun$ to the secondary star, assuming it is accreted or deflected by winds.

\begin{figure}
 \centering
 \includegraphics[]{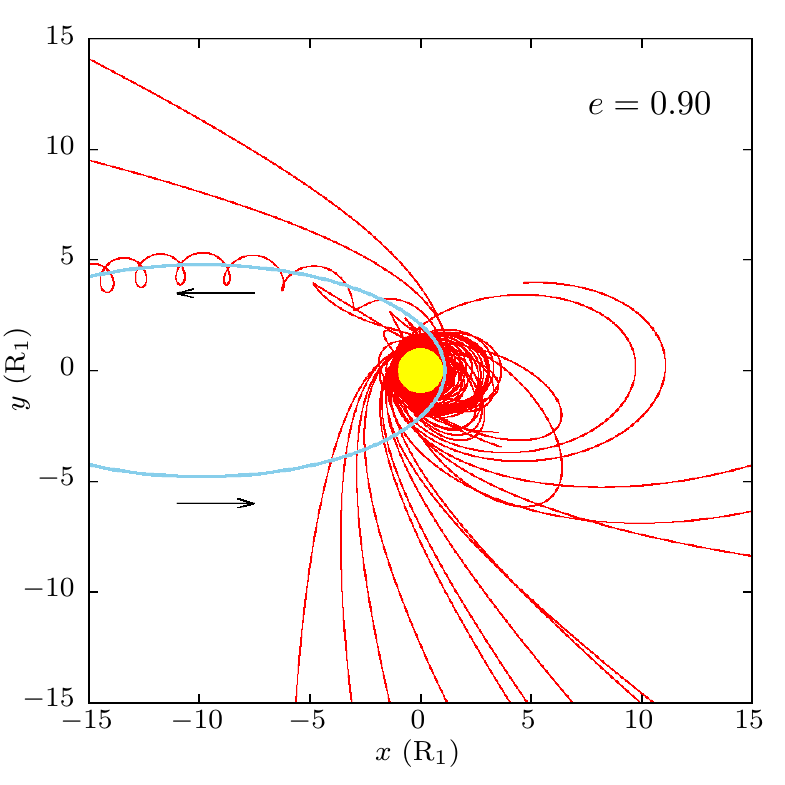}
 \caption{Trajectories of the test particles around an eccentric binary in the frame of the primary star (red curves). The yellow circle represents the primary star radius and the light blue ellipse indicates the trajectory of the secondary star. Simulation parameters are $q=0.5, e=0.9$, $a(1-e)=1.1~R_1$ and results are shown for only one binary orbit.\label{fig:spray_trajectory_e09}}
\end{figure}

\begin{figure}
 \centering
 \includegraphics[]{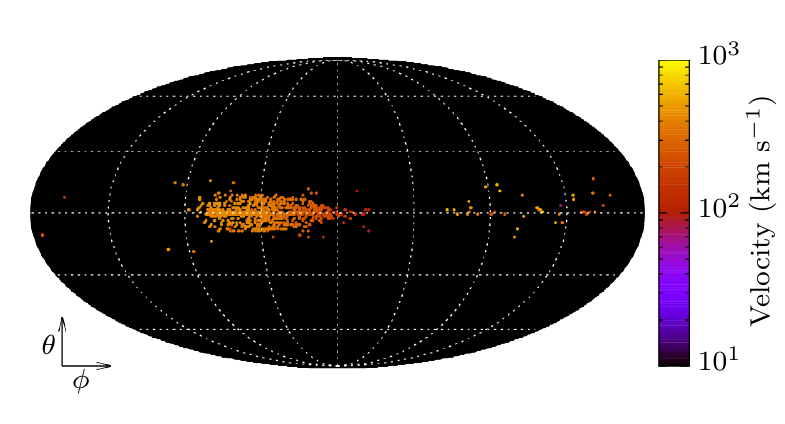}
 \caption{Mollweide projection of the test particle ejection directions in the same simulation as Figure~\ref{fig:spray_trajectory_e09}. The origin is taken as the direction of the primary to secondary at periastron (eccentricity vector) and the span of $\varphi$ at $\theta=0$ defines the orbital plane. $\varphi$ increases in the direction of the orbit. Colours of plots show the ejection velocity for $R_1=20\rsun$ and $M_1=60\msun$.\label{fig:mollweide}}
\end{figure}

Figure~\ref{fig:spray_trajectory_e09} shows that test-particle trajectories behave in roughly three different ways. Some are simply lifted off the surface slightly and placed on eccentric bound orbits around the primary. A second group wrap around the trajectory of the secondary star, indicating mass transfer. But a third group reaches out of the panel, becoming unbound from the system and so forming spray ejecta. The sky map of spray directions is shown in Figure~\ref{fig:mollweide}. Velocities are calculated by asigning a scale of $R_1=20~\rsun$ and $M_1=60~\msun$. Most spray particles are confined within $\lesssim15\degree$ from the orbital plane. There is a clear cluster of spray particles on the left which are the particles ejected downwards in Figure~\ref{fig:spray_trajectory_e09}. Because these particles are already weakly bound to the star due to the spin, the gravitational lift at periastron sends them on hyperbolic orbits around the primary and out of the system. There is another weak cluster on the right which corresponds to the particles shooting out to the left in Figure~\ref{fig:spray_trajectory_e09} and are more spread out. These particles are flung away from the close vicinity of the secondary star. 

\begin{figure}
 \centering
 \includegraphics[]{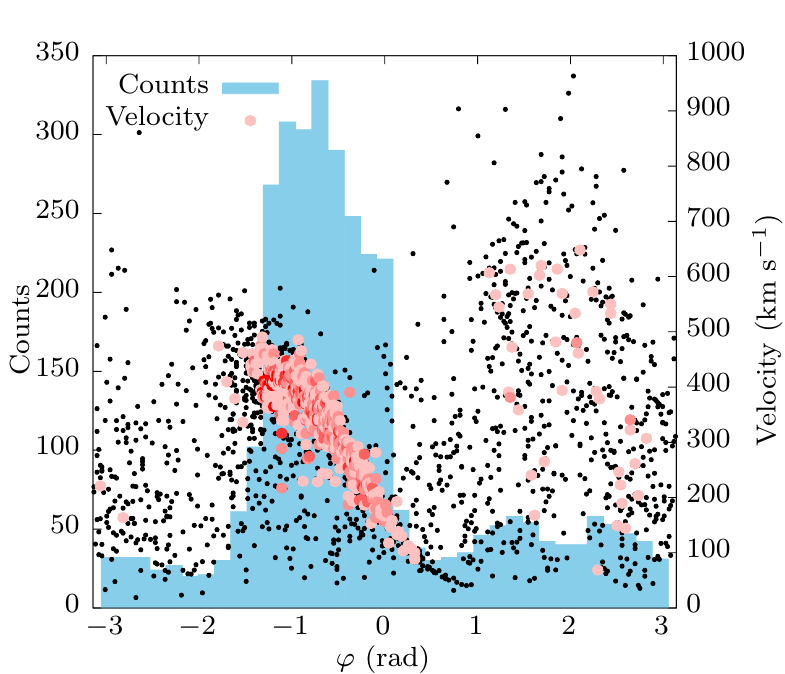}
 \caption{Histogram of the test particle ejection directions in the same simulation as in Figures \ref{fig:spray_trajectory_e09} \& \ref{fig:mollweide}. Overplotted are the velocity of spray particles in each $(\theta,\varphi)$ bin. Pink and red dots show the average velocity of particles where there were multiple counts in the same angular bin where the redder dots have more counts. Black dots are the velocity where there was only one count in the bin.\label{fig:eje_histogram}}
\end{figure}

We then split up the sky into $(N_\theta\times N_\varphi)=(180\times360)$ bins and calculate the average spray velocity in each direction. Figure~\ref{fig:eje_histogram} shows the velocity in each bin overplotted on a histogram of the number of particles ejected in each $\varphi$ direction. It is clear that the majority of the ejected particles are clustered in one direction with an opening angle of $\sim90\degree$. The sprays have velocities in the range 100--500~km~s$^{-1}$ while an extremely small number of particles reach up to $\sim900~$km~s$^{-1}$. The maximum velocity of the bulk ejecta ($\sim500$~km~s$^{-1}$) is roughly determined by the asymptotic velocity of a hyperbolic orbit which was launched at the surface of the primary and had an initial angular velocity equivalent to the orbital angular velocity at periastron
\begin{align}
 v_\mathrm{max}=\sqrt{\frac{G(M_1+M_2)(1+e)}{a(1-e)}\cdot\frac{R_1}{a(1-e)}-\frac{2GM_1}{R_1}}.\label{eq:vmax}
\end{align}
The velocity roughly follows a linear distribution, with maximum velocity at $\varphi\sim-\pi/2$ and declining up to $\varphi\sim$0.

\begin{figure}
 \centering
 \includegraphics[]{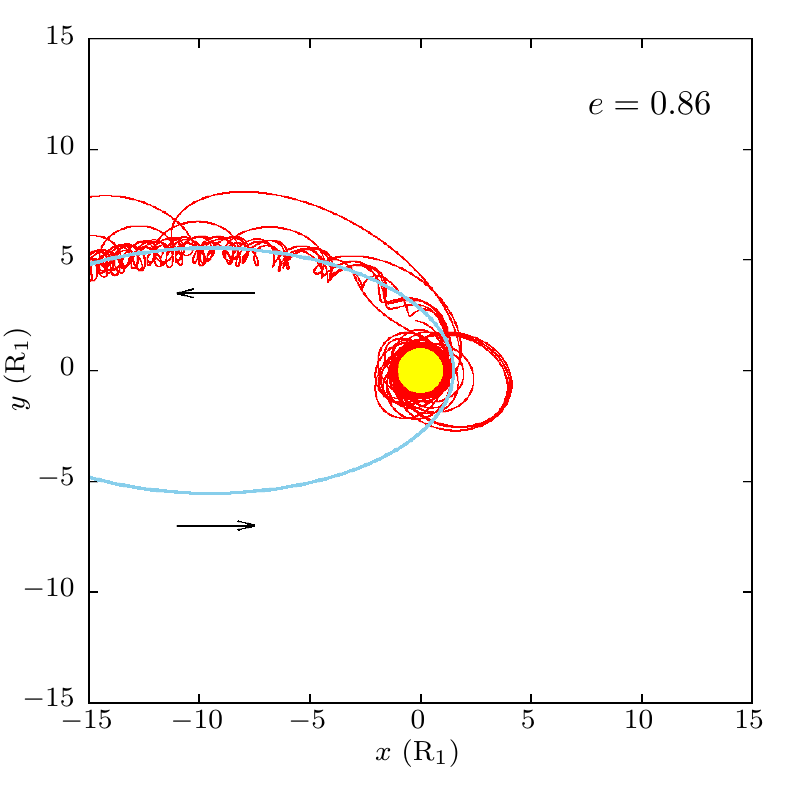}
 \caption{Same as Figure~\ref{fig:spray_trajectory_e09} but with slightly lower eccentricity. Other parameters are chosen to keep the semimajor axis fixed. Simulation parameters are $q=0.5, e=19/22\sim0.86$ and $a(1-e)=1.5~R_1$.\label{fig:spray_trajectory_e08}}
\end{figure}

\begin{figure}
 \centering
 \includegraphics[]{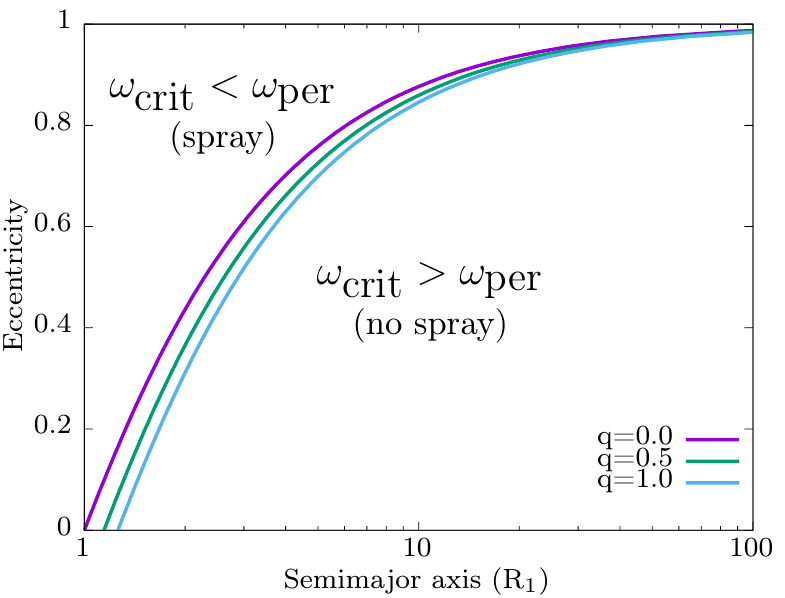}
 \caption{Critical eccentricities where the orbital angular velocity at periastron equals the critical spin angular velocity of the primary star. Three different curves are shown for different mass ratios. The horizontal axis is in units of primary star radius.\label{fig:spray_criterion}}
\end{figure}

To explore how the sprays depend on the orbital parameters, we have carried out an additional set of simulations. We find that when the semimajor axis is fixed, just a slight decrease in eccentricity can significantly affect the amount of spray ejecta. Figure~\ref{fig:spray_trajectory_e08} shows the particle trajectories for the case where $e=19/22$. There are a few particles being transferred to the secondary but none of them are ejected from the system. For eccentricities of $e=0.8$ we found that there are not even any transferred particles. A comparison between the three different eccentricities can be seen in the movie available online (Movie~4). It seems that there is a well-defined threshold above which sprays can happen. This can be understood by comparing the orbital angular velocity at periastron 
\begin{equation}
 \omega_\mathrm{per}= \sqrt{\frac{G(M_1+M_2)(1+e)}{a^3(1-e)^3}},
\end{equation}
 and the critical spin angular velocity 
\begin{equation}
 \omega_\mathrm{crit}=\sqrt{\frac{GM_1}{R_1^3}}.
\end{equation}
When $\omega_\mathrm{crit}<\omega_\mathrm{per}$, the secondary star exhibits a spin-up torque to the primary. If the surface is already rotating at critical, these particles are marginally bound so the small acceleration can make them unbound, resulting in spray ejection. In other cases, the secondary exhibits a spin-down torque, so the surface particles never become unbound. Figure~\ref{fig:spray_criterion} shows when the two angular velocities cross over. In a hierarchical triple system, the Kozai-Lidov mechanism causes the inner orbit to librate and change eccentricity while the semimajor axis is fixed. Note that at any given semimajor axis, there exists a critical eccentricity $e_\mathrm{crit}$ above which spray ejection can occur. For the examples shown in Figures \ref{fig:spray_trajectory_e09} \& \ref{fig:spray_trajectory_e08}, the semimajor axis is fixed to $a/R_1=11$ and the transition between spraying and non-spraying cases can be well explained by this critical eccentricity. This is of course strongly tied to the assumption that the primary is already rapidly spinning. This is probably not a bad assumption since the primary star in this case is the mass accretor in the previous evolution. In our scenario, the mass-transfer stage is not too long before the spray ejection stage, so any spin-down mechanism would not have actively spun down the star yet. Also, the tidal torques from the orbit at periastron will keep it rapidly rotating.

The above simulation indicates that eccentric close encounters can send out material in confined directions. There is a preferred direction perpendicular to the eccentricity vector, with an opening angle of $\sim90\degree$. Along with the ejection velocity ($\sim$100--500~km~s$^{-1}$), these periastron sprays are in good agreement with the observed sprays \citep[]{kim16}. Multiple periastron passages will send out several sprays, which will interact with each other. Self-interaction shocks could quickly cool the material, and make them clumpy. We have only tracked the motion of test particles, so we are not able to give quantitative estimates on the amount of ejected mass. There is very little mass located in the surface layers. For example, there is only $\sim$0.1$~\msun$ in the outer $\sim$1/3 of the stellar radius for a 50$~\msun$ main sequence star. Therefore the total amount of ejected mass through these close encounters should be considerably less https://www.overleaf.com/project/5f445c31ef47a700014dedf1than $\lesssim$0.1$~\msun$. It should also be noted that we have only considered purely gravitational effects in this simple calculation. In reality the stars could graze each others surface, and non-gravitational effects such as shocks and radiation would influence the mass ejection from the system too. Further research is required to investigate how much the other interactions can contribute to the spray ejecta and whether they have similar ejection directions, and how much mass can be ejected.

Because the sprays are produced from the surface material of the mass accretor (star B), it is possible that they have chemical peculiarities. Indeed, the observed spray material are known to be nitrogen rich \citep[]{dav82}, with a decreasing amount of N enrichment farther from the star \citep{sm04}. It will be interesting to estimate the precise chemical composition of the spray ejecta in future studies.

\subsection{Spatial distribution of sprays}
By combining our simulation results in the previous sections, we can predict the spatial distribution of spray ejecta around the Homunculus. For simplicity, we do not carry out a large grid of spray simulations but instead fit a simple functional form to the velocity distribution of spray particles
\begin{equation}
 v_\mathrm{spray}=\left(-\frac{\varphi}{\pi}+\frac{1}{2}\right)v_\mathrm{max} \quad\left(-\frac{\pi}{2}<\varphi<0\right),\label{eq:spray_dist}
\end{equation}
where $v_\mathrm{max}$ is computed from Eq.~(\ref{eq:vmax}). The forefactor simply accounts for the linear distribution in $\varphi$. Although not a perfect fit to the spray simulation results, this roughly gives the velocity scaling and angular distribution (pink and red dots) in Figure~\ref{fig:eje_histogram}. 

\begin{figure*}
 \centering
 \includegraphics[]{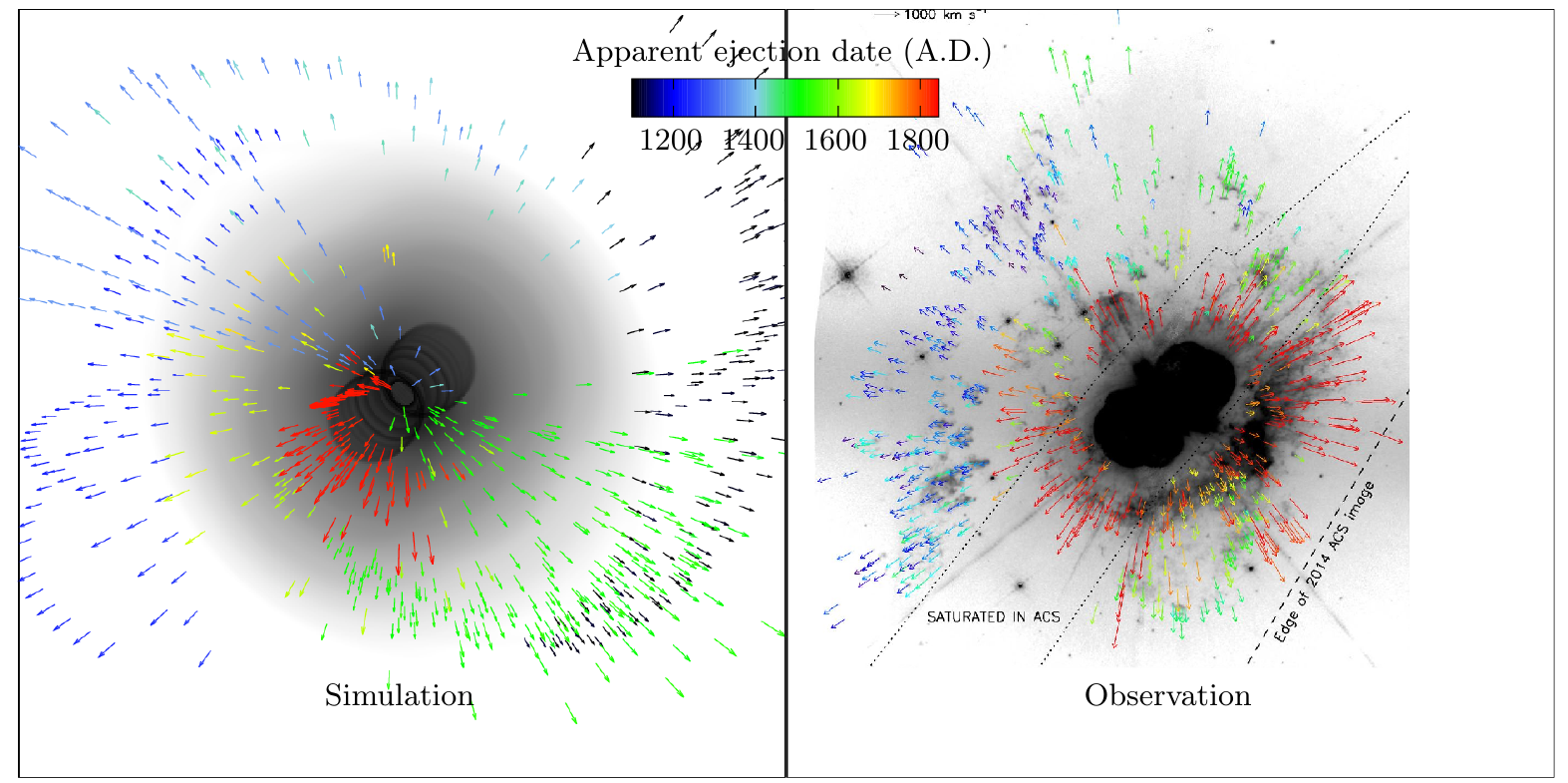}
 \caption{(\textit{left panel}) Spatial distribution of sprays in our simulation at 176~yr after the Great Eruption. Colours of vectors indicate the apparent ejection date assuming the merger occured in 1844. The lengths of vectors are proportional to its velocity. The vectors are overlaid on a mock H$\alpha$ emission measure image based on our hydrodynamical simulation. A movie showing the time evolution and 3D distribution of sprays is available online (Movie~5). (\textit{right panel}) Observed distribution of Outer Ejecta directly taken from Figure~7 of \citet{kim16}.\label{fig:altogether}}
\end{figure*}

We next apply this formula to the sample triple system in section \ref{sec:triple}. We assume that stars B and C started merging when the distance between the two stars first reached down to $20~\rsun$ ($t\sim1972$~yr). At every periastron passage before that, where $e>e_\mathrm{crit}$, we assume that a spray was ejected along the orbital plane with a velocity distribution following Eq.~(\ref{eq:spray_dist}). By assuming the spray ejecta follow ballistic trajectories, we can calculate the positions of ejecta at arbitrary times. In Figure~\ref{fig:altogether} we show the position and velocity vectors of spray ejectiles 176~yr after the Great Eruption. Here we have assumed that the Great Eruption occurred 30~yr after the onset of the merger. To mimick the clumpiness of the sprays, we have only selected 3--4 random directions within the opening angle for each periastron passage. The particles are colour coded by the apparent ejection date assuming the Great Eruption occurred in 1844. A Homunculus model is placed at the centre, with the symmetry axis taken in the direction of the orbital angular momentum vector of the merging binary. There are 5--6 distinct ejection episodes in this figure, corresponding to the eccentricity peaks seen in Figure~\ref{fig:distance_timeevo}. Each ejection is one-sided and points in a different direction. None of the ejections are aligned with the symmetry axis of the Homunculus or with the equatorial plane. This highly asymmetric and randomly directed nature is remarkably consistent with the observed Outer Ejecta \citep[right panel;][]{kim16}. The velocities also show an increasing trend towards more recently ejected sprays. This is because Eq.~(\ref{eq:vmax}) gives higher velocities for higher eccentricities, and the eccentricity increases over time in this system (Figure~\ref{fig:distance_timeevo}). The closer-in red vectors could correspond to the ghost shell or outer shell observed in H$\beta$ \citep[]{meh16}. The background image is produced by integrating $\rho^2$ along each line of sight, representing the H$\alpha$ emission.

We have not considered any overtaking in this calculation. In reality the faster later sprays that overtake slower earlier sprays can wipe out their traces, reducing the number of sprays observed today and possibly enhance the clumpiness. Also, we have made an arbitrary choice for the delay time between the onset of merger and the Great Eruption. We can, however, constrain the delay time to some degrewe. \citet{kim16} suggest that the prior mass ejections occurred in 1250, 1550 and early 1800s. If the 1800s outburst was due to a close encounter, the merger should have happened after 1800 but before the Great Eruption in 1844, meaning that the spiral-in process took less than 44 years. We have also suggested that the precursor eruptions in 1838 and 1843 were due to interactions between the binary companion and the bloated  envelope during the spiral-in phase (Figure~\ref{fig:framework}, Panel 3-2). This indicates that the common-envelope phase started before 1838, placing a lower limit of $\sim6$~yr for the spiral-in time-scale. So the merger delay time can be narrowed down to $\sim$6--50~yr, which will not significantly affect the results shown in Figure~\ref{fig:altogether}.

The example demonstrated in this section is not the only possible path. For example, the companion swap could have happened much earlier and the sprays could have occurred through interactions between stars B and C. There could have been a triple common-envelope phase where the envelope of the tertiary expands to engulf the inner binary \citep[]{gla21}. By the time the envelope is shedded, the tertiary becomes a hydrogen-poor star and the inner and outer orbits can both shrink to an unstable configuration. The sprays and merger can occur shortly after. Our basic picture does not change in any case. The main point is that the system starts off as a hierarchical triple system but becomes unstable once one of the stars evolves off the main sequence. Chaotic Kozai-Lidov cycles enables close encounters of stars and sends out spray ejecta. The unstable orbit eventually triggers a merger which causes the Great Eruption and shapes the bipolar Homunculus nebula.

The one-sided ejections, random orientations and $\sim300$~yr intervals of the sprays are all difficult to explain in most of the previously proposed models. Whereas in the triple evolution scenario proposed by \citet{smi18b} that we investigate in more detail here, all these features are naturally expected. From the above experiments, we now confirm that the historical eruptions and Outer Ejecta are no longer counter-arguments to the merger scenario but may instead be supporting evidence for the model.

\section{Post-merger evolution}\label{sec:postmerger}

Following the merger, we have assumed that the energy from the merger that was not carried away in the eruption is radiated as enhanced luminosity, observed initially to range up to a value $2.5 \times 10^7 L_\odot$ \citep{fre04,smi11a}. This is well above the Eddington luminosity, and so can drive a mass loss up to the energy limit \citep{owo17}, which here can be more than two orders of magnitude higher than even the very strong current-day mass loss of $10^{-3}~\msun$~yr$^{-1}$. The strong wind should cease once all the excess energy is radiated away and the star regains thermal equilibrium. Although the post-merger star could have a non-standard chemical profile due to the strong mixing during the merger, the star should continue its evolution as a normal single star of the mass of the merger product \citep[]{sch20}. If any of the merging stars was depleted in hydrogen in the core, it could experience a prolonged phase of hydrogen shell burning, spending more time as a blue supergiant than that of a star of the same mass \citep[]{gle13}.

Stars in this mass range ($\sim100~\msun$) are expected to have very strong winds and experience luminous blue variable activity, so the mass-loss rate should be relatively high even without the enhanced luminosity due to the merger. Therefore {\etac} can lose nearly a half or more of its own mass during its lifetime. The final fate of these stars are rather uncertain. One possible outcome is that it simply collapses into a black hole. In this case, most of the mass in the star will be retained and could form a several$\times10~\msun$ black hole. The kick velocity imparted to black holes are expected to be quite small, so the binary would not disrupt. Because the likely outcome of the companion star ({\etac} B) is also black hole formation, the system could become a very eccentric binary black hole. The other possible fate of the merger product is a pulsational or non-pulsational pair-instability supernova. In the former case, it will experience several large pulsations and end up as a normal core-collapse or failed supernova. In the latter case the whole star will be expelled in the explosion and will not leave any remnant. However, this may be unlikely at solar metallicities ($Z>Z_\odot/3$) due to the large mass loss that hinders the establishment of a massive enough core \citep[]{lan07}.

We also note that, if the merger takes place with the original primary (i.e.\ if star A is not swapped out of the inner orbit), which is already a post-main-sequence star at the time of the merger, the subsequent evolution of the merger product could be quite different \citep[][see also \citealt{van13,vig19}]{jus14}. Specifically, it is likely to spend most of its helium core-burning phase as a blue supergiant and become a luminous blue variable (of the S Doradus type) shortly before it explodes in a supernova \citep[]{jus14}.

\section{Discussion}\label{sec:other}
In our eruption simulation we have not taken into account any other possible sources of energy other than the gravitational and orbital energy of the core binary. Other processes may tap in extra sources of energy such as explosive nuclear burning when fresh fuel is dragged in to the higher temperature regions \citep[]{iva02a,iva02b}. However, this would only be significant if one of the merging stars have depleted hydrogen in their core, which is not the case for our fiducial model laid out in Figure~\ref{fig:framework}. Magnetic fields could also play an important role \citep[]{sch19}. We have also only used a single model for our combined mass ($100~\msun$) but the total mass could have been larger. In that case the core masses would have been larger too, implying that there was more orbital energy in the core binary. All these effects could tap in more energy to the eruption, leading to more violent explosions and therefore more ejecta mass. These processes may be fundamentally important if the true Homunculus mass is much greater than the inferred lower limits \citep[$\gtrsim$10--45$~\msun$;][]{smi03b,mor17}.

The highest velocities achieved in our eruption simulations reached up to $\sim$10,000~km~s$^{-1}$ although the mass in that high velocity matter is tiny. Such velocities are in fact observed in light echoes \citep[]{smi18a}. It should be noted that, in our simulation, the equatorial shock breakout has slightly faster velocities than the polar shock breakout. This is likely caused by the way we set the background density. Because we set a slow wind as a background, the density is higher at smaller radii. Therefore the density contrast between the stellar surface and outside is smaller around the poles, so the immediate shock breakout velocity is slightly slower. Because the mass in the fast components are small, it will easily be decelerated as it runs into slower pre-eruption wind material that was flowing at 150--200~km~s$^{-1}$ \citep[]{smi18b}. The degree of deceleration will depend on the mass-loss rate of the pre-eruption wind. At some point this fast Outer Ejecta material will catch up and collide with even denser pre-eruption ejecta such as the spray material from the historical ejections or L2 outflow material from the mass-transfer phase. It has been proposed that this collision between extremely fast ejecta from the Great Eruption and slower pre-eruption ejecta is the origin of the observed soft X-ray shell around $\eta$~Carinae \citep{sm04,smi08}. X-rays are in fact observed from the outer regions and the estimated shock velocities are 700--800~km~s$^{-1}$ \citep[]{sew01,wei04}. This is roughly consistent with the relative velocity between spray ejecta velocities and the eruption ejecta velocity inferred from the location of the X-rays\footnote{The eruption ejecta velocity is simply $v=r/t$ where $r$ is the distance from the star and $t$ is the time since the Great Eruption.}.

This scenario does not naturally explain the origin of the lesser eruptions that occurred after the Great Eruption. The lesser eruption around 1890 is considered to have created the Little Homunculus that lies inside the Homunculus nebula itself \citep[]{ish03,smi05}. Its inferred apparent ejection date is around 1910--1930, meaning that if it was really ejected in the 1890s, the Little Homunculus is likely being swept up and accelerated in a similar way to the main Homunculus in our scenario \citep[]{smi05}. What caused the 1890s eruption is still an open question. Some studies claim that the lesser eruptions were triggered by interactions with the secondary at periastron \citep[]{kas10}. However, their model requires extremely high masses for both the primary and secondary stars ($M_1\sim200~\msun$, $M_2\sim80~\msun$), which are factors of 2--3 larger than the observationally inferred values.

Some other important visible features of the Outer Ejecta are the ``NN jet'', S condensation and the equatorial ``skirt'' \citep[]{wal76,mea96,mor98,kim16,meh16}. These structures have slightly older apparent ages, but could be consistent with being ejected in the Great Eruption if it was decelerated later on \citep{mor01,kim16,smi17}. Moreover, the NN jet and the S condensation seem to be aligned with the direction of some non-axisymmetric features of the Homunculus known as ``protrusions'' \citep[]{ste14}. The protrusions are located roughly $\sim$110$\degree$ apart, where the centre points in the direction of the opening in the CO torus observed by ALMA \citep[]{smi18c}. Together with the fact that the apocentre of the current-day binary orbit points towards the gap in the torus \citep[]{mad12}, \citet{smi18b} proposed that all these structures were possibly shaped by the binary companion plunging through the bloated common envelope or circumstellar torus. In Phase 3-2 of our scenario (see Fig.~\ref{fig:framework}), the companion could have plunged through the envelope multiple times, punching a hole in it. The momentum of the companion wind is relatively small, so it could only drill small tunnels. As pointed out by \citet{smi18a}, the later Great Eruption will be channeled through the holes, squirting out a narrow feature like the NN jet. However, it is not clear whether the envelope distortions could have been sustained until the Great Eruption since the rotational period of the envelope is relatively short ($\sim1$~yr) and the rotation could have quickly smeared out any small distortions.

\section{Conclusion}\label{sec:summary}
We have performed a suite of numerical simulations to investigate the merger-in-a-triple scenario for the origin of $\eta$ Carinae and its multiple eruptions similar to the model proposed by \citet{smi18b} \citep[see also][]{fit12}. Our study confirms that this scenario gives a plausible explanation for the Great Eruption of $\eta$ Carinae. In addition, our simulations suggest that the strong bipolar wind from the merger product played a critical role in sweeping up and shaping the bipolar Homunculus after the eruption.

We first carry out 2.5D hydrodynamical simulations of the outflow from an explosive stellar merger event. The ejecta follow a homologous expansion, and are distributed in a rather spherical manner. We then inject a strong bipolar wind following a standard gravity-darkening law to see how the ejecta get swept up into a thin shell. We find that we can reproduce the shape of the Homunculus nebula fairly well, although with some remaining questions about the latitudinal and radial mass distribution.

Because {\etac} is a binary system today, the system must have started off as a triple system if {\etac} is indeed a merger product. We expect that the system was unstable in the past and could have experienced swaps of companions and close encounters of stars which eventually triggered the merger. We demonstrate through 3-body dynamical simulations that an unstable triple system can induce companion swaps and close encounters under the right conditions. We also carry out N-body simulations of how the surface material on a star reacts to the periastron passage in a highly eccentric orbit. When the eccentricity exceeds a certain threshold, the surface particles can be ejected in one direction with a $\sim90\degree$ opening angle along the orbital plane. By combining the triple dynamical evolution and the mass ejection from single close encounters, we estimate how the ejected matter should be distributed around the Homunculus. The one-sided nature of each ejection, the seemingly random ejection directions and $\sim100$~yr intervals all agree well with the observed distribution of the Outer Ejecta \citep{kim16}.

Both our hydrodynamical simulations of mass outflow from the merger and dynamical simulations of the pre-merger evolution reproduce many of the key features of $\eta$ Car's observed nebula, strongly supporting the merger-in-a-triple scenario \citep{smi18b}. We can therefore use $\eta$ Car as a prototype to deepen our understanding of stellar mergers in general. Massive stellar mergers can be responsible for a diverse range of phenomena, such as luminous blue variables and supernova impostors \citep{smi11c}, peculiar supernovae like SN1987A \citep{pod90}, magnetic stars \citep{sch19}, or creating interaction-powered supernova progenitors \citep{jus14,smi14}. The B[e] star binary R4 in the Small Magellanic Cloud is a particularly similar case that has been claimed to be the outcome of a stellar merger in a triple system \citep{pas00,wu20}. Given the relatively high fraction of massive stars in triple systems \citep[]{rag10,moe17}, it may mean that \etac-like systems are not uncommon. The rich observational data for $\eta$ Car can therefore not only help us decipher its own history, but also the origin of many other important astrophysical phenomena.

\section*{Acknowledgements}
The authors thank the anonymous referee for the constructive comments that improved the content. The authors thank Lorne Nelson for sharing computational facilities. The computations were partially carried out on facilities managed by Calcul Qu\'ebec and Compute Canada. RH was supported by the JSPS Overseas Research Fellowship No.29-514 and a grant from the Hayakawa Satio Fund awarded by the Astronomical Society of Japan. The work has also been supported by a Humboldt Research Award to PhP. at the University of Bonn. SPO acknowledges a Royal Society International Exchange grant that supported his visit to the University of Oxford in the early stages of this project. FRNS has received funding from the European Research Council (ERC) under the European Union’s Horizon 2020 research and innovation programme (Grant agreement No. 945806). NS received support from National Science Foundation (NSF) grant AST-1515559, and from NASA grants GO-14586, GO-15289, GO-15596, and GO-15823 from the Space Telescope Science Institute, which is operated by the Association of Universities for Research in Astronomy, Inc. under NASA contract NAS 5-26555.

\section*{Data Availability}
The data underlying this article will be shared on reasonable request to the corresponding author.

\bibliographystyle{mnras}

%%%%%%%%%%%%%%%%% APPENDICES %%%%%%%%%%%%%%%%%%%%%

\appendix

\section{Basic equations and Code description for hydrodynamical simulation}\label{app:code}

To simulate the dynamical process of our scenario, we solve the Euler equations for hydrodynamics. We assume axial symmetry and use a spherical coordinate system ($r,\theta,\varphi$), in which the continuity equation becomes
\begin{eqnarray}
% \frac{\partial\rho}{\partial t}+\nabla\cdot(\rho\bm{v})=0,\\
 \frac{\partial\rho}{\partial t}+\frac{1}{r^2}\frac{\partial}{\partial r}(r^2\rho v_r)+\frac{1}{r}\frac{\partial}{\partial\theta}(\sin\theta\rho v_\theta)=0,
\end{eqnarray}
where $\rho$ is density, $t$ is time, and $v_r, v_\theta$ are the radial and polar components of velocity ($\bm{v}$) respectively, defined on physical bases. Similarly the equations of motion can be written in conservative form as
\begin{eqnarray}
 \frac{\partial(\rho v_r)}{\partial t}+\frac{1}{r^2}\frac{\partial}{\partial r}\left(r^2(\rho v_r^2+p)\right)+\frac{1}{r}\frac{\partial}{\partial\theta}\left(\sin\theta\rho v_rv_\theta\right)\nonumber\\
=\frac{\rho(v_\theta^2+v_\varphi^2)+2p}{r}-\rho\frac{\partial\phi}{\partial r},\\
 \frac{\partial(\rho v_\theta)}{\partial t}+\frac{1}{r^2}\frac{\partial}{\partial r}(r^2\rho v_rv_\theta)+\frac{1}{r}\frac{\partial}{\partial\theta}\left(\sin\theta(\rho v_\theta^2+p)\right)\nonumber\\
=-\frac{\rho v_rv_\theta}{r}+\frac{\rho v_\varphi^2+p}{r}\cot\theta-\frac{\rho}{r}\frac{\partial\phi}{\partial\theta},\\
 \frac{\partial(\rho v_\varphi)}{\partial t}+\frac{1}{r^2}\frac{\partial}{\partial r}(r^2\rho v_rv_\varphi)+\frac{1}{r}\frac{\partial}{\partial\theta}\left(\sin\theta\rho v_\theta v_\varphi\right)\nonumber\\
=-\frac{\rho v_\theta v_\varphi}{r}\cot\theta-\frac{\rho v_rv_\varphi}{r}+\rho r\sin\theta\dot{\Omega}_\mathrm{add},
\end{eqnarray}
where $p$ is pressure, $\phi$ the gravitational potential and $v_\varphi$ is the azimuthal component of the velocity. $\dot{\Omega}_\mathrm{add}$ is an additional source for angular velocity which is explained in the main text. Although we assume axial symmetry, we still take into account rotational velocities so the vector components contain azimuthal components. This kind of approach is sometimes called a 2.5-dimensional (2.5D) approach. The energy conservation equation is
\begin{eqnarray}
 \frac{\partial e}{\partial t}+\frac{1}{r^2}\frac{\partial}{\partial r}\left(r^2(e+p)v_r\right)+\frac{1}{r}\frac{\partial}{\partial\theta}\left(\sin\theta(e+p)v_\theta\right)\nonumber\\
=-\rho\nabla\phi\cdot\bm{v}+\rho r\sin\theta\dot{\Omega}_\mathrm{add}v_\varphi,
\end{eqnarray}
where $e$ is the sum of kinetic and internal energy. The gravitational field $\phi$ is obtained through the Poisson equation
\begin{eqnarray}
 \Delta\phi=4\pi G\rho.
\end{eqnarray}
For the eruption simulation, we assume that the whole star is optically thick and is in local thermal equilibrium. Then the temperature $T$ can be defined through 
\begin{eqnarray}
 \epsilon=\frac{3}{2}\frac{k_\mathrm{B}T}{\mu m_\mathrm{u}}+\frac{a_\mathrm{rad}T^4}{\rho},
\end{eqnarray}
where $\epsilon$ is specific internal energy, $k_\mathrm{B}$ is the Boltzmann constant, $\mu$ the mean molecular weight, $m_\mathrm{u}$ the atomic mass unit and $a_\mathrm{rad}$ the radiation constant. The equation of state includes contributions from gas and radiation
\begin{eqnarray}
 p=\frac{\rho k_\mathrm{B}T}{\mu m_\mathrm{u}}+\frac{a_\mathrm{rad}T^4}{3}.
\end{eqnarray}
 For the sweep-up simulation, we assume that the unbound ejecta has become optically thin enough for radiation to decouple from the gas. Therefore the material is not in local thermal equilibrium any more so we drop the radiation pressure term in the equation of state. Radiative cooling also becomes efficient at this point so we incorporate this by using the scheme proposed in \citet{tow09}. For cooling efficiencies we use a piecewise power-law fit to the tabulated collisional ionization equilibrium values in \citet{gna07}\footnote{http://wise-obs.tau.ac.il/~orlyg/cooling/}. We apply a floor temperature $T_\mathrm{floor}=10^4$~K for the cooling because there is no information for the cooling efficiency below this. Physically, this can be interpreted as being heated by radiation from the central star.

To solve these equations we use the hydrodynamical code \textsc{hormone} (High ORder Magneto-hydrodynamic cOde with Numerous Enhancements) developed by one of us \citep[]{RH16}. This is a grid-based code that solves the ideal magneto-hydrodynamic equations through a Godunov type scheme. Second-order spatial accuracy is acquired based on MUSCL interpolation with a generalized minmod flux limiter \citep[]{van79}. The conserved variables are interpolated and pressure is re-calculated at the interface using the interpolated values. Whenever the pressure gradient is inverted because of the interpolation, we switch to first-order reconstruction to avoid spurious motions. The fluxes across cell boundaries are calculated by the HLLD approximate Riemann solver \citep[]{miy05}, which is equivalent to the HLLC solver when magnetic fields are neglected. A dimensionally unsplit, total-variation-diminishing Runge-Kutta scheme of 3rd order is used for time integration \citep[]{shu88}. The code has been updated so that the advection of chemical elements can be taken into account using the consistent multi-fluid advection method \citep[]{ple99}. Outgoing boundary conditions are applied to the outer boundary. For self-gravity we evolve the gravitational field using the hyperbolic self-gravity method with a gravitation propagation speed factor $k_\mathrm{g}=10$ \citep[see][]{RH16} but capped by the speed of light. We apply a Robin boundary condition for the outer boundary condition of self-gravity \citep[]{gus98}. 

\begin{figure}
 \centering
 \includegraphics[]{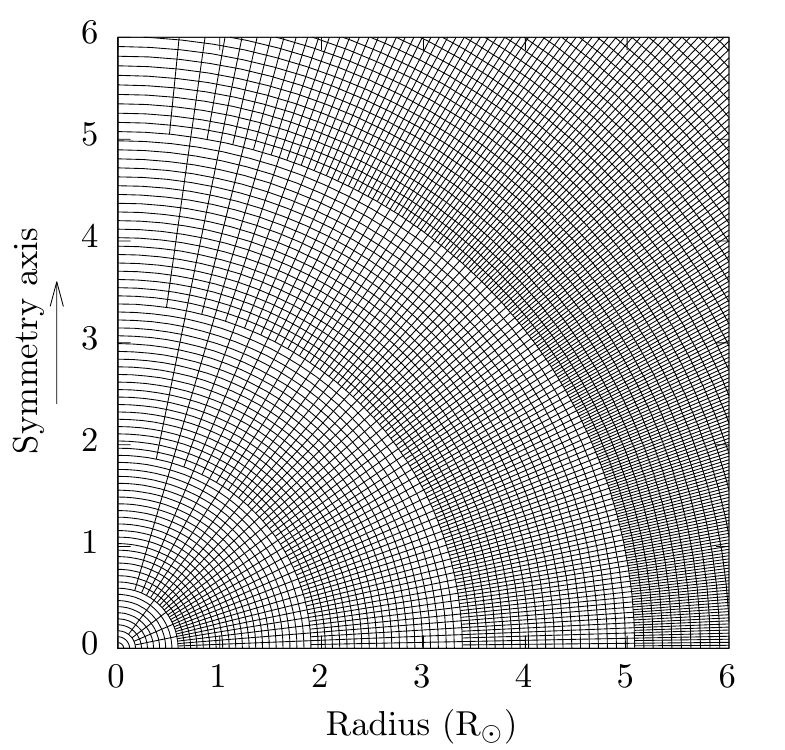}
 \caption{Effective grid structure of the 2.5D hydrodynamical simulations.\label{fig:grid}}
\end{figure}

Cell sizes are increased as a geometrical series in the radial direction, and uniform spacings in $\cos\theta$ are used for the polar direction. We use $N_r=900$ and $N_r=1700$ for the eruption and sweep-up simulations, respectively. For the eruption simulation, the radial size of the innermost cell is $\Delta r_0=0.06~\rsun$ and the outer boundary is set at 3000~$\rsun$, so that the entire star and ejecta is well contained within the computational domain throughout the simulation. For the central regions where the Courant conditions are most severe, we effectively reduce the number of polar grid points by averaging conserved quantities over a few cells after each time step. Vector quantities are averaged in a way that the total angular momentum is conserved. The central region is effectively 1D for the first 3 radial cells and then $N_\theta=5$ for the next 7, $N_\theta=25$ for the next 20, $N_\theta=50$ for the next 20, $N_\theta=100$ for the next 20 and $N_\theta=200$ for the rest of the grid. Our effective grid structure can be seen in Figure~\ref{fig:grid}. We only apply this grid structure to the hydrodynamics and not to the self-gravity solver. We check that the star is stable for several years on this grid in the absence of angular momentum and energy injection. 

For the sweep-up simulation, we change the radial computational domain adaptively to efficiently follow the motion of the shell. 

\section{Estimating the shell position}\label{app:shell}
The radial density distribution of the ejecta in each latitudinal direction roughly resembles that of 1\,D eruption simulations \citep[]{owo19}. The inner parts of the density distribution along each latitudinal direction can be fitted by
\begin{eqnarray}
 \rho_\mathrm{ej}(r,t)=\frac{\Delta M}{8\pi v_0^3t^3}e^{-r/v_0t},\label{eq:densityfit}
\end{eqnarray}
where $\Delta M$ is the spherical equivalent mass in that direction and $v_0$ is a fitting parameter that expresses the steepness of the distribution. The velocity distribution is linear $v_\mathrm{ej}=r/t$ which is characteristic of a homologous expansion.

\begin{figure}
 \centering
 \includegraphics[]{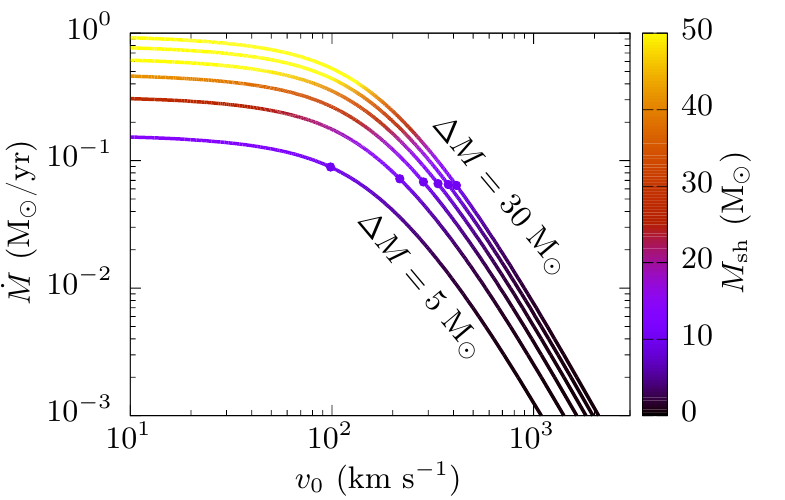}
 \caption{Wind mass-loss rate required to achieve $v_\mathrm{sh}=650$~km~s$^{-1}$ wind for various ejecta profiles ($\Delta M, v_0$). The wind velocity is assumed to be $v_w=1000$~km~s$^{-1}$. Each curve corresponds to different $\Delta M$ values in the range $\Delta M=5$--$30~\msun$, with $5~\msun$ intervals. Colours of the curves display the total mass contained in the thin shell. The purple circles indicate where the shell mass is $M_\mathrm{sh}=10~\msun$.\label{fig:windshellrelation}}
\end{figure}

In the sweep-up simulation we inject a stellar wind from the inner edge of this ejecta. Although we do not know the outcome of the simulation a priori, we can estimate the final shell velocity in the following way. Assuming spherical symmetry and that at time $t$ the wind has swept up the inner ejecta into a thin shell up to $r_\mathrm{sh}$ at a velocity $v_\mathrm{sh}$, we can write down the total momentum balance as
\begin{align}
 &4\pi\int^{r_\mathrm{sh}}_0\rho_\mathrm{ej}\frac{r}{t}r^2dr+\dot{M}v_wt=\nonumber\\
&\left[4\pi\int^{r_\mathrm{sh}}_0\rho_\mathrm{ej}r^2dr+\dot{M}\left(t-\frac{r_\mathrm{sh}}{v_w}\right)\right]v_\mathrm{sh}+\dot{M}v_w\cdot\frac{r_\mathrm{sh}}{v_w},\label{eq:analytic_shell}
\end{align}
where $\dot{M}$ is the wind mass-loss rate and $v_w$ is the wind velocity. The first term on the left hand side simply integrates the momentum in the ejecta and the second term is the momentum added by the wind, so the sum represents the total momentum available up to $r_\mathrm{sh}$. The first term on the right hand side represents the momentum that is in the thin shell and the second term is the momentum in the wind that has not reached the shell yet. Assuming that the shell is moving at roughly the same speed as the ejecta immediately above, it gives a relation $r_\mathrm{sh}=\eta v_\mathrm{sh}t$ where $\eta$ is a parameter close to but smaller than unity. By applying the density fitting function as given in Eq.~\ref{eq:densityfit}, the integrals can be calculated analytically
\begin{eqnarray}
 \frac{1}{2}\Delta Mv_0f\left(\eta,\frac{v_\mathrm{sh}}{v_0}\right)+\dot{M}t\left[v_\mathrm{sh}\left(1+\eta-\eta\frac{v_\mathrm{sh}}{v_w}\right)-v_w\right]=0,
\end{eqnarray}
where
\begin{align}
 f(\eta,x)=&\left[\eta^2(\eta-1)x^3+\eta(3\eta-2)x^2+(6\eta-2)x+6\right]e^{-\eta x}\nonumber\\
&+2x-6.
\end{align}
The equation gives a relation between wind properties ($\dot{M}, v_w$) and shell properties ($r_\mathrm{sh}, v_\mathrm{sh}$), so we can obtain the wind parameters required to produce the observed shells by plugging in observed values and solving for $\dot{M}$. The mass in the shell can also be calculated by
\begin{equation}
 M_\mathrm{sh}=\frac{1}{2}\Delta Mg\left(\frac{\eta v_\mathrm{sh}}{v_0}\right)+\dot{M}t\left(1-\frac{\eta v_\mathrm{sh}}{v_w}\right),
\end{equation}
where
\begin{equation}
 g(x)=(x^2+2x+2)e^{-x}-2.
\end{equation}
The first term on the right hand side represents the mass from the ejecta material swept up into the shell and the second term represents the contribution of mass added by the wind. We display some of the relations in Figure~\ref{fig:windshellrelation}, plugging in the observed shell and wind velocities at the pole \citep[$v_\mathrm{sh}=650$~km~s$^{-1}$ and $v_w=1000$~km~s$^{-1}$, e.g.][]{smi06a}, and assuming the shell is moving at exactly the same velocity as the ejecta immediately above it ($\eta=1$). For example, for an ejecta profile with $\Delta M=10~\msun$ and $v_0=200$~km~s$^{-1}$, a wind mass-loss rate of $\dot{M}\sim10^{-1}~\msun$~yr$^{-1}$ is required to reproduce the current day shell velocity. This is two orders of magnitude larger than the current observed mass-loss rate of $\dot{M}\sim10^{-3}~\msun$~yr$^{-1}$. However, it is likely that the wind mass-loss rate was significantly larger after the Great Eruption and decayed over time. The high mass-loss rate only needs to be sustained for the first $180~\mathrm{yr}\cdot(1-v_\mathrm{sh}/v_w)\sim60$~yr because any later wind has not reached the shell yet and does not affect the current day shell properties. If we further constrain the relation with the observed shell mass \citep[$M_\mathrm{sh}\gtrsim10~\msun$,]{smi06a}, the required mass-loss rate is $\dot{M}\gtrsim7\times10^{-2}~\msun$~yr$^{-1}$, and seems weakly dependent on the ejecta profiles (purple circles in Figure~\ref{fig:windshellrelation}).

This is just a simple way to estimate the shell velocity by sorting out the total available momentum into wind, shell and ejecta. Many assumptions are made in this analysis such as spherical symmetry, infinitesimally thin shell, or the shell moving at the same speed as the incident ejecta ($\eta=1$). A more rigorous way to estimate the shell properties through ejecta-wind interaction will be presented in a subsequent paper (Owocki et al. in prep.)

\section{Drag force within radiative envelopes}\label{app:drag}

It is often assumed that stars will merge when the two stars touch each others surface. However, this is not necessarily the case for stars with radiative envelopes because the density in these envelopes are extremely low. When compact stars such as WR stars plunge through such envelopes, the drag force that acts on the star can be negligible compared to its orbital energy.

\begin{figure}
 \centering
 \includegraphics[]{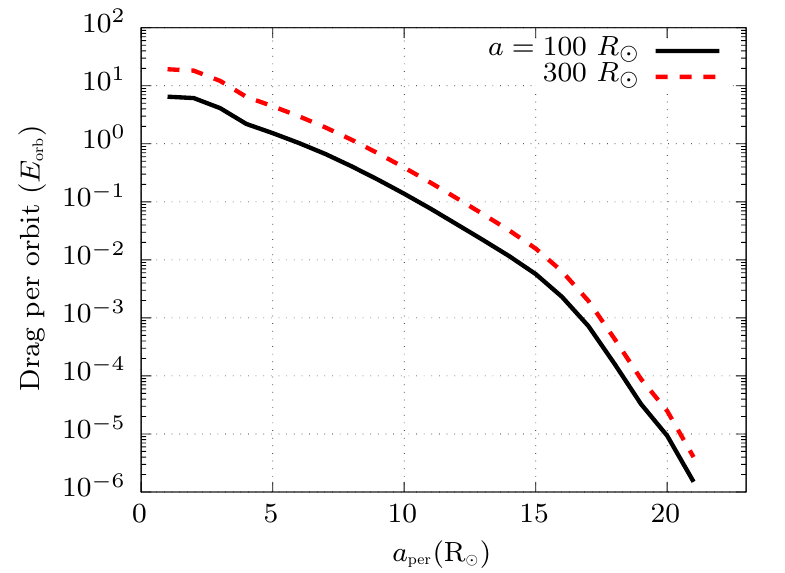}
 \caption{Drag per orbit from the interaction of a WR star plunging into a 50~$\msun$ main-sequence star. The horizontal axis is the periastron distance, which is a measure of the eccentricity. Each curve is calculated with different semimajor axes for the orbit.\label{fig:drag_in_star}}
\end{figure}

Here we carry out a simple calculation to quantify this effect. We assume that a $M_2=30~\msun$ WR star is orbiting a $M_1=50~\msun$ main-sequence star. This is meant to represent the inner binary in the system we simulated in Section \ref{sec:triple}. In the triple evolution towards merger, the semimajor axis of the inner orbit is roughly constant until it reaches the chaotic stage at the very end. The eccentricity modulates due to Kozai-Lidov oscillations and sometimes reaches very high values. When the eccentricity is high, the periastron distance is small enough for the companion to plunge into the envelope of the primary. The drag force acting on the WR star can be estimated by 
\begin{equation}
 F_\textrm{drag}=\frac{1}{2}\rho v_\textrm{rel}^2A,
\end{equation}
where $v_\textrm{rel}$ is the relative velocity of the two bodies and 
\begin{equation}
 A=\pi\left(\frac{2GM_2}{v_\textrm{rel}^2}\right)^2
\end{equation}
is the cross-sectional area within the Bondi radius of the WR star. By integrating this force over the orbit, we can calculate how much energy the drag takes away upon each plunge. In Figure~\ref{fig:drag_in_star} we plot the drag per orbit as a function of periastron distance. Note that the drag is normalized by the orbital energy $E_\textrm{orb}$, so it shows what fraction of the orbital energy is taken away per revolution. The radius of this star is $\sim22~\rsun$, but even if the companion plunges in as deep as $\sim15~\rsun$, the orbit is only affected by $\sim1~\%$ so it will require hundreds of orbits to have a significant effect.

The eccentricity peaks in Figure~\ref{fig:distance_timeevo} only last for a couple of orbits, so it is fair to say that any interaction between star A and the envelope of star B will not affect the evolution of the triple significantly. Also, the mass ejected in the grazing encounters will be very small because of the low density in the outer layers of these radiative envelopes.

\bsp
\label{lastpage}
\end{document}